\documentclass[twocolumn]{aastex62}

\usepackage{scrextend}
\usepackage{hyperref}

\graphicspath{{./}{figures/}}

\submitjournal{ApJ}

\shorttitle{Properties of Y-Dwarfs}
\shortauthors{Zalesky et al.}

\begin{document}

\title{A Uniform Retrieval Analysis of Ultracool Dwarfs. III: Properties of Y-Dwarfs}

\correspondingauthor{Joseph A. Zalesky}
\email{joezalesky@asu.edu}

\author[0000-0001-9828-1848]{Joseph A. Zalesky}
\affil{School of Earth and Space Exploration, Arizona State University, Tempe, AZ 85287, USA}

\author[0000-0002-2338-476X]{Michael R. Line}
\affil{School of Earth and Space Exploration, Arizona State University, Tempe, AZ 85287, USA}

\author[0000-0002-6294-5937]{Adam C. Schneider}
\affil{School of Earth and Space Exploration, Arizona State University, Tempe, AZ 85287, USA}

\author{Jennifer Patience}
\affil{School of Earth and Space Exploration, Arizona State University, Tempe, AZ 85287, USA}


\begin{abstract}

Ultra-cool brown dwarfs offer a unique window into understanding substellar atmospheric physics and chemistry. Their strong molecular absorption bands at infrared wavelengths, Jupiter-like radii, cool temperatures, and lack of complicating stellar irradiation, make them ideal test-beds for understanding Jovian-like atmospheres. Here we report the findings of a uniform atmospheric retrieval analysis on a set of 14 Y and T-dwarfs observed with the Hubble Space Telescope Wide Field Camera 3 instrument. From our retrieval analysis, we find the temperature-structures to be largely consistent with radiative-convective equilibrium in most objects. We also determine the abundances of water, methane, and ammonia and upper limits on the alkali metals sodium and potassium. The constraints on water and methane are consistent with predictions from chemical equilibrium models, while those of ammonia may be affected by vertical disequilibrium mixing, consistent with previous works. Our key result stems from the constraints on the alkali metal abundances where we find their continued depletion with decreasing effective temperature, consistent with the trend identified in a previous retrieval analysis on a sample of slightly warmer late T-dwarfs in \citet[]{2017ApJ...848...83L}. These constraints show that the previously observed Y-J color trend across the T/Y transition is most likely due to the depletion of these metals in accordance with predictions from equilibrium condensate rainout chemistry. Finally, we simulate future James Webb Space Telescope observations of ultra-cool dwarfs and find that the NIRSpec PRISM offers the best chance at developing high-precision constraints on fundamental atmospheric characteristics. \\

\end{abstract}

\section{Introduction} \label{sec:intro}

	Brown dwarfs have solicited intriguing questions since their discovery several decades ago \citep{1988Natur.336..656B,1995Natur.377..129R, 1995Sci...270.1478O}. While not being massive enough to to fuse hydrogen into helium \citep{1963PThPh..30..460H, 1963ApJ...137.1121K}, they were still too massive to be considered as ``traditional" planets following the roughly 13M$_{\rm Jup}$ definition based on the fusion of deuterium \citep{1977ApJ...214..488S,1996ApJ...460..993S}. More recently there have been arguments that formation pathways, rather than mass limits, are more useful when defining the difference between brown dwarfs and planets \citep{2001ApJ...551L.167B,2003MNRAS.339..577B}. This has placed the study of brown dwarfs at an interesting crossroads between planetary science and stellar astrophysics. Efforts to understand the physics of brown dwarfs have thus pulled methodologies from both fields in order to measure the physical characteristics and understand the evolution of these objects \citep[for a review,][]{2015ARA&A..53..279M}.
    
    Motivation for studying the atmospheres of brown dwarfs is two-fold. First, brown dwarfs do not have a stable internal energy source, and thus their evolution is highly dependent upon their initial formation mass \citep[e.g.][]{2003A&A...402..701B} and specific physical/chemical structure of their atmosphere \citep[e.g.][]{2008ApJ...689.1327S}. Secondly, brown dwarfs offer the chance to study planetary-like atmospheric conditions, while not having to include the complication of an irradiating host star. Understanding the physical and chemical mechanisms at work in cooler brown dwarf atmospheres thus provides constraints on both their evolution, and the characteristics of planetary-like atmospheres.
    
    The bulk properties of field brown dwarfs (mass, radius, T$_{\rm eff}$, etc.) have been well studied over the past several decades \citep[for a review,][]{2001RvMP...73..719B}. With cool effective temperatures ($200K \lesssim$ T$_{\rm eff}$ $\lesssim 3000K$) over photospheric pressures ($300\lesssim$ P $\lesssim 10^{-4}$ bar), their thermal emission predominately radiates in the near-to-mid infrared, with their spectra being sculpted by strong molecular and atomic opacities of species such as: hydrogen and helium (H$_2$/He), water (H$_2$O), methane (CH$_4$), ammonia (NH$_3$), and alkali metals such as potassium (K) and sodium (Na) for the coolest objects to carbon monoxide and dioxide (CO,CO$_2$), H$_2$O, H$_2$/He, and metal hydrides and oxides for the hottest \citep{1996ApJ...472L..37F, 2002Icar..155..393L, 2003ApJ...591.1220L}. The precise molecular and cloud compositions, and their evolution with temperature, give rise to spectral signatures which define the L-T-Y spectroscopic classes \citep{1995Sci...270.1478O, 1999ApJ...519..802K, 2005ApJ...623.1115C, kirkpatrick2005, 2011ApJ...743...50C}.
    
	While empirical approaches exist \citep[e.g.][]{2009AJ....137.3345C,2015ApJ...810..158F}, the primary method of choice for inferring atmospheric properties relies upon detailed comparisons between theoretical models and the observed spectra. This often takes the form of pre-computing a large grid of theoretical spectra across a range of key physical parameters \citep{1996ApJ...465L.123A, 2012RSPTA.370.2765A, 1996Sci...272.1919M, 1996A&A...308L..29T}. Most commonly these grids include effective temperature and gravity, but more recently have been modified to include variable cloud models \citep{2001ApJ...556..872A, 2002ApJ...568..335M}, eddy diffusion within the atmosphere \citep{2006ApJ...647..552S}, rainout of specific condensates, and varying metallicity and carbon-to-oxygen (C/O) ratios \citep{2017AAS...23031507M, 2017A&A...600A..10M, 2017A&A...603A..57S}. These large grid models are then interpolated between grid points and fit via standard maximum likelihood comparisons \citep[e.g.][]{2008ApJ...678.1372C} or modern MCMC methods \citep[e.g.][]{2011ApJ...729...41M,2015PASP..127..479R,2015ApJ...804...64M,2017A&A...603A..57S}.
    
    Though this grid modeling approach provides a useful baseline for beginning the analysis of infrared spectra, it has been shown to fail to accurately reproduce key spectral features, and often provides poor fits to the data \citep[e.g.][]{2017ApJ...842..118L}. For example, \citet{2012A&A...540A..85P} has demonstrated that grid models from different groups cannot reproduce statistically similar results for the same dataset of young brown dwarf companions. These inconsistencies between grid model fitting and the observational data suggest that not all of the possible atmospheric physics and chemistry is being taken into account within the established grid models. Despite this, a more recent effort in \citet{2017ApJ...850..150B} found greater consistency between several widely-used grid models, though outstanding issues in abundance determinations still remain. These inconsistencies motivate the need for a new methodology to compliment the grid-modeling approach to reach a more complete understanding of brown dwarf spectra.
    
    Realizing the limitations of the grid modeling approach,  \citet{2015ApJ...807..183L, 2017ApJ...848...83L} (hereafter Parts I \& II) applied well established atmospheric retrieval \citep{twomey1977, 2007Icar..189..457F, 2012MNRAS.420..170L, 2012ApJ...749...93L, 2012ApJ...753..100B} tools to the problem by performing a uniform retrieval analysis on a sample of late-T dwarfs. In Part I, the authors were able to validate their model on two benchmark T-dwarfs by showing that the overall retrieved abundances and C/O ratios were consistent with the objects' stellar companion. With the larger sample (11 T7-T8 targets) in Part II, they found a strong depletion of the combined Na+K abundances with decreasing T$_{\rm eff}$. This had long been a theoretical expectation from rainout chemistry \citep{1994Icar..110..117F}, and hypothesized from trends of near-infrared colors \citep{2002ApJ...568..335M,2010ApJ...710.1627L, 2012ApJ...758...57L, 2013A&A...550L...2L}, but the measured abundance depletion had never been directly detected. These investigations demonstrate that the retrieval method as applied to brown dwarf atmospheres is able to constrain key atmospheric properties often overlooked in traditional methods.
    
    Our primary goals in this work, Part III, are to both expand the previously analyzed dataset into the cooler, early-Y dwarf (Y0-Y1) regime to see if the trends identified in Part II continue to cooler temperatures, and to test the various model assumptions made in Parts I \& II. This is accomplished by performing a retrieval analysis on a set of objects from \citet{2015ApJ...804...92S}, which contains near-IR (0.9-1.7$\mu m$, Y,J,H band) spectra of 6 late-T and 16 early-Y dwarfs using Hubble Space Telescope's Wide Field Camera 3 (HST,WFC3).
    
    In Section \ref{sec:methods} we briefly outline the methods of our atmospheric retrieval model. Section \ref{sec:dataset} discusses the dataset from WFC3 and the history of our targets. Constraints on the temperature structure (Section \ref{sec:temp}), evolutionary parameters (Section \ref{sec:gravity}), and chemical abundances (Section \ref{sec:comp}) are then discussed. We also perform a comparison of our retrieval method with a recently published grid model in Section \ref{sec:grid}. In Section \ref{sec:jwst} we predict how well the future \textit{James Webb Space Telescope} (\textit{JWST}) will be able to improve our constraints. Finally, we list our primary conclusions in Section \ref{sec:conclusions}.
    
\begin{table}
\renewcommand{\thetable}{\arabic{table}}
\centering
\caption{31 Free-Parameters in Our Retrieval Model} \label{tab:mod_params}
\begin{tabular}{cc}
\tablewidth{0pt}
\hline
\hline
Parameter & Description \\
\hline
\decimals
log(f$_i$) & 7 log(constant-with-altitude \\ 
& Volume Mixing Ratios) \\
log(g) & log(GM/R$^2$) [cm s$^{-2}$] \\
(R/D)$^2$ & radius-to-distance scale [R$_{\rm Jup}$/pc]\\
T(P) & temperature at 15 pressure levels [K] \\
$\Delta \lambda$ & wavelength calibration uncertainty [nm] \\
$b$ & errorbar inflation exponent \\
& (Part I, Equation 3) \\
$\gamma, \beta$ & TP-profile smoothing hyperparameters\\
& (Part I, Table 2/Equation 5)\\
$\kappa_{P_0}, P_0, \alpha$ & Cloud profile parameters\\
& (Part II, Equation 1)\\
\hline
\end{tabular}
\end{table}

\section{Methods} \label{sec:methods}

    \begin{table*}
\renewcommand{\thetable}{\arabic{table}}
\centering
\caption{Basic photometric properties of our sample.} \label{tab:obs}
\begin{tabular}{cccccc}
\tablewidth{0pt}
\hline
\hline
\decimals
WISE/AllWISE Name & Spec. Type & Y$_{\rm MKO}$\footnote{\label{adamphot}Synthetic Photometry from \citet{2015ApJ...804...92S}. } [mag] & J$_{\rm MKO}$\footref{adamphot} [mag] & H$_{\rm MKO}$\footref{adamphot} [mag] & Dist. [pc]\\
\hline
WISEA J032504.52-504403.0 & T8 & 19.980$\pm$0.027 & 18.935$\pm$0.024 & 19.423$\pm$0.027 & 27.2$\pm$2.2\footnote{\label{kirk2019dist}Distances from \citet{2019ApJS..240...19K}}. \\
WISEA J040443.50-642030.0 & T9 & 20.328$\pm$0.032 & 19.647$\pm$0.025 & 19.970$\pm$0.033 & 21.9$\pm$1.4\footref{kirk2019dist} \\
WISEA J221216.27-693121.6 & T9 & 20.282$\pm$0.023 & 19.737$\pm$0.024 & 20.225$\pm$0.036 & 12.2$\pm$0.4\footref{kirk2019dist}\\
WISEA J033515.07+431044.7 & T9 & 20.166$\pm$0.029 & 19.467$\pm$0.023 & 19.938$\pm$0.031 & 13.9$\pm$0.5\footref{kirk2019dist}\\
WISEA J094306.00+360723.3 & T9.5 & ... & 19.766$\pm$0.025 & 20.315$\pm$0.038 & 10.7$\pm$0.3\footref{kirk2019dist}\\
WISEA J154214.00+223005.2 & T9.5 & 20.461$\pm$0.028 & 19.937$\pm$0.026 & 20.520$\pm$0.045 & 11.6$\pm$0.6\footref{kirk2019dist}\\
WISEA J041022.75+150247.9 & Y0 & ... & 19.325$\pm$0.024 & 19.897$\pm$0.038 & 6.52$\pm$0.17\footnote{\label{martindist}Distances from \citet{2018ApJ...867..109M}.}\\
WISEA J073444.03-715743.8 & Y0 & 20.870$\pm$0.041 & 20.354$\pm$0.029 & 21.069$\pm$0.071 & 15.1$\pm$1.2\footref{martindist}\\
WISEA J173835.52+273258.8 & Y0 & ... & 19.546$\pm$0.023 & 20.246$\pm$0.031 & 7.34$\pm$0.22\footref{martindist}\\
WISEA J205628.88+145953.6 & Y0 & ... & 19.129$\pm$0.022 & 19.643$\pm$0.026 & 7.23$\pm$0.20\footref{martindist}\\
WISEA J222055.34-362817.5 & Y0 & 20.899$\pm$0.034 & 20.447$\pm$0.025 & 20.858$\pm$0.035 & 11.9$\pm$0.75\footref{martindist}\\
WISEA J163940.84-684739.4 & Y0 Pec. & 20.833$\pm$0.023 & 20.626$\pm$0.023 & 20.764$\pm$0.029 & 4.39$\pm$0.10\footref{martindist}\\
WISEA J140518.32+553421.3 & Y0.5 & 21.33$\pm$0.057 & 21.061$\pm$0.035 & 21.501$\pm$0.073 & 6.76$\pm$0.49\footref{martindist}\\
WISE J154151.65-225024.9 & Y1 & 20.461$\pm$0.028 & 19.934$\pm$0.026 & 20.520$\pm$0.045 & 5.98$\pm$0.14\footref{martindist}\\
\hline
\end{tabular}
\end{table*}

    We utilize the same basic retrieval framework and forward model as described in Part's I and II, briefly summarized here. The model includes 31 free-parameters which include: constant-with-altitude volume mixing ratios (VMR's) for: H$_2$O, CH$_4$, NH$_3$, CO, CO$_2$, H$_2$S, and a combined alkali [Na+K] fixed at a solar ratio (7 in total), gravity, radius-to-distance scale factor $(R/D)^2$, 15 independent temperature-pressure (TP) profile points implemented within a Gaussian-Process-like smoothing framework, and a simple cloud parameterization \citep{2006ApJ...650.1140B}; summarized in Table \ref{tab:mod_params}. All of our molecular and alkali cross-sections are those of \citet{2008ApJS..174..504F, 2014ApJS..214...25F}. We have also implemented new Na and K cross-sections from \citet{2016A&A...589A..21A} but found no substantial change to our retrieved abundances.
   
    One aspect to re-iterate is that we neglect scattering, which may break down in the presence of strongly forward-scattering clouds. Part II did not find any strong evidence for the presence of optically thick clouds in the late-T dwarf sample, as expected for cooler brown dwarfs \citep[e.g.][]{kirkpatrick2005}, though the alkali depletion trend was consistent with the expected trend in the Na$_2$S and KCl condensation-temperature profile intersections. Y-dwarfs are cooler than late T-dwarfs, permitting the possible formation of water-ice clouds at low pressures \citep{2014ApJ...787...78M}, though we note that the presence of water clouds would have minimal impact over the 1.0-1.7$\mu m$ wavelength range covered by WFC3 \citep[see Figure 16b of ][]{2018arXiv180407771M}. 
    
	At the WFC3 spectral resolution (R$\sim$140) and signal-to-noise ratio (J-band SNR$\sim$20), the alkalis' overall spectral signal, if present, is blended together to create a continuum-like absorption feature along the red portion of the Y-band due to the broad wings of the 0.59 and 0.77$\mu$m resonance lines, and weak features at 1.24$\mu m$ in the J-band. As was done in Parts I \& II, we have combined their signatures and kept the [Na/K] fixed to the solar ratio for this reason. We have experimented with relaxing this assumption and allowing both Na and K to be retrieved independently, but find no substantial difference in our results.
    
    To solve the parameter estimation problem we use the affine-invariant MCMC sampler \verb|emcee| \citep{2013PASP..125..306F}, initialized using a grid-model profile TP profile from \citet{1996Sci...272.1919M} interpolated to our 15 level parameterization. We choose an approximate T$_{\rm eff}$ and gravity for each object based on the spectral type from \citet{2015ApJ...804...92S} and approximate thermochemical equilibrium abundances for a representative pressure.  As in Parts I \& II, we have  checked that our MCMC chains have converged (typically 40-60K iterations with 200 walkers) and that our results are not sensitive to the initial starting conditions for our model.

\section{Dataset} \label{sec:dataset}

In both Parts I \& II, ground-based near-IR spectra from the SpeX Prism Library \citep{2014ASInC..11....7B} were used. As our aim is to extend into the cooler Y dwarf regime, we turn to space-based spectroscopy in order to have comparable SNR on cooler targets. Our chosen dataset are the 6 late-T and 16 early-Y objects observed in \citet{2015ApJ...804...92S} with HST's WFC3, which details the WFC3 data reduction process. This sample was chosen as it provides the most complete, uniformly reduced spectra of the known Y dwarfs.

For all of our targets we have used the most recent distance estimates available in the literature \citep{2018ApJ...867..109M,2019ApJS..240...19K}. We have also done an analysis assuming parallax estimates from several other authors, the results of which are presented in Section \ref{sec:gravity} \citep{2012ApJS..201...19D, 2016AJ....152...78L, 2017MNRAS.468.3764S}.

We first performed an initial retrieval on all 22 of our targets. While the retrieval technique obtains constraints on various atmospheric parameters, it is ultimately a data-driven technique which requires precise spectroscopy to properly converge. We found that 8 of our 22 targets had low enough SNR to prevent our retrieval model from converging upon physically-realistic TP profiles, and thus have not included them in our analysis.

A brief review the remaining 14 objects, specified by their Wide-field Infrared Survey Explorer (WISE) identification, is provided below. Key observational quantities (YJH magnitudes and distance estimates) are summarized in Table \ref{tab:obs}.
    
    \noindent
    \textbf{WISEA J032504.52-504403.0 (W0325, T8):} This is one of several new brown dwarf discoveries from \citet{2015ApJ...804...92S}. The object spectroscopically well-matches the T8 spectral standard \citep{2006ApJ...637.1067B}. Follow-up work was done by \citet{2017ApJ...842..118L} who published an archival J band magnitude. \\
    \noindent
    \textbf{WISEA J040443.50-642030.0 (W0404, T9):}  This is another new discovery from \citet{2015ApJ...804...92S} who found the spectrum to closely match the T9 spectral standard \citep{2011ApJ...743...50C}.\\
    \noindent
    \textbf{WISEA J221216.27-693121.6 (W2212, T9):} Another new discovery of \citet{2015ApJ...804...92S} which is also in good agreement with the T9 spectral standard.\\
    \textbf{WISEA J033515.07+431044.7 (W0335, T9):}  This object was discovered in the \citet{2013ApJS..205....6M} study along with 86 other T dwarfs and given the classification of T9. Both \citet{2014ApJ...783...68B} and \citet{2015ApJ...804...92S} found good agreement with this classification. \citet{2015ApJ...799...37L} published new ground-based YJHK photometry from Gemini Observatory for this target and noted an unusually faint K-band measurement with respect to their other T dwarfs. They found that current equilibrium models could not well-explain their photometric measurements unless the assumed NH$_3$ abundance was halved and/or there were systematic issues with the CH$_4$ linelist. \citet{2017ApJ...842..118L} used archival photometry combined with non-equilibrium models from \citet{2015ApJ...804L..17T} to conclude that this target may have sub-solar metallicity.\\
    \textbf{WISEA J094306.00+360723.3 (W0943, T9.5):} W0943 was initially discovered in \citet{2014AJ....147..113C} who published WISE and Hubble photometry along with WFC3 G141 spectroscopy. It was given a classification of T9.5 which agree well with the results from \citet{2015ApJ...804...92S}.\\
    \textbf{WISEA J154214.00+223005.2 (W1542, T9.5):} W1542 was also discovered in the \citet{2013ApJS..205....6M} study and given the classification of T9.5 which is in agreement with other similar studies \citep{2014ApJ...783...68B,2015ApJ...804...92S}. \\
    \noindent
    \textbf{WISEA J041022.75+150247.9 (W0410, Y0):}  This is a well studied object that was part of the initial classification of Y dwarfs as distinct from their T dwarf counterparts \citep{2011ApJ...743...50C}. These measurements helped establish the extreme blue shift of Y-J colors across the T/Y transition \citep[e.g.][]{2012ApJ...758...57L, 2012ApJ...753..156K}. \citet{2014AJ....147..113C} obtained the first space-based spectrum of this object and confirmed Y0 classification. \citet{2013ApJ...763..130L} found that fitting YJHK photometry from the Near-Infrared Imager on Gemini North with cloudy models from \citet{2012ApJ...756..172M} results in effective temperatures, gravities, and low cloud sedimentation efficiencies consistent with previous analyses of early-Y dwarfs. Spectra from \citet{2015ApJ...804...92S} agree well with the photometry of \citet{2013ApJ...763..130L}. \citet{2015ApJ...799...37L} highlighted that current equilibrium models were unable to reproduce their updated photometry for early-Y dwarfs and that retrieval techniques are needed to understand these objects. \citet{2017ApJ...842..118L} visually fit the spectra of \citet{2015ApJ...804...92S} with cloud-free, vertical disequilibrium models from \citet{2015ApJ...804L..17T} and found that while Y and H band are visually well-fit, the J-band model spectrum was $\sim$20\% brighter.\\
    \noindent
    \textbf{WISEA J073444.03-715743.8 (W0734, Y0):} This is another object initially discovered by an early WISE Survey \citep{2012ApJ...753..156K}. Follow-up photometric work by both \citet{2014ApJ...796...39T} and \citet{2015ApJ...799...37L} noted the rather red Y-J color for this object, more consistent with a late-T than an early-Y dwarf. \\
    \noindent
    \textbf{WISEA J173835.52+273258.8 (W1738, Y0):} This object is probably one of the most well-studied cool brown dwarfs as it represents the Y0 spectral standard \citep{2011ApJ...743...50C}. \citet{2012ApJ...750...74S} introduced updated collisionally induced H$_2$ and NH$_3$ opacities and found improved fits to observed infrared colors. \citet{2013A&A...550L...2L} provided Z-band imaging of this and several other cool brown dwarfs. \citet{2013ApJ...763..130L} noted the rather blue Y-J colors now seen in many Y dwarfs. \citet{2015MNRAS.448.3775R} performed the first photometric monitoring of such a cool target, however this object proved too faint to provide the precision needed to confirm variability in the J-band. Recent ground-based spectra have revealed that non-equilibrium models from \citet{2015ApJ...804L..17T} better-fit the entire near-infrared spectrum of this object, however a majority of the Y-band is not well fit \citep{2016ApJ...824....2L}. \citet{2016ApJ...830..141L} found a peak-to-peak 3\% Spitzer [4.5] variability consistent with W1738's rotation period of roughly 6hrs.\\
    \noindent
    \textbf{WISEA J205628.88+145953.6 (W2056, Y0):} W2056 represents another archetypal WISE early-Y dwarf also analyzed in the \citet{2011ApJ...743...50C} and \citet{2012ApJ...753..156K} studies. \citet{2013ApJ...763..130L} obtained ground-based YJHK photometry and far-red spectra for this object. They noted that overall, cloudy models from \citet{2012ApJ...756..172M} fit both the red spectra and K band well but were too faint near 1.0, 1.5, and 1.65$\mu m$. This was attributed to both overly-strong NH$_3$ absorption due to vertical mixing of NH$_3$ not being included in the models, and incomplete CH$_4$ molecular line lists. \citet{2017ApJ...842..118L} provided new ground-based M' observations for this object, and visually fit cloud-free models from \citet{2015ApJ...804L..17T} to archival WFC3 spectra. Due to poor S/N and sparse sampling in their grid model, the archival WFC3 spectra were fit only by-eye.\\
    \noindent
    \textbf{WISEA J222055.34-362817.5 (W2220, Y0):} W2220 was initially discovered in the \citet{2012ApJ...753..156K} study as a new Y dwarf. Chosen as part of a astrometric survey, \citet{2014ApJ...783...68B} noted that W2220 provided tentative evidence for variability due to more than a magnitude difference between archival J and H-band measurements. \citet{2017ApJ...842..118L} concluded this object is consistent with a solar-metallicity and solar-age field dwarf. \\
    \noindent
    \textbf{WISEA J163940.84-684739.4 (W1639, Y0 Pec.):} W1639 was first discovered by \citet{2012ApJ...759...60T} after carefully resolving the near-infrared counterpart to the WISE point source with ground-based imagining. Though J-band spectroscopy matches well to the Y0 standard, both Y-band spectra and Y-J colors deviate from the standard, leading to the Y0-Peculiar classification \citep{2015ApJ...804...92S}. \citet{2016ApJ...819...17O} searched for, but found no evidence of another companion to the known Y dwarf to within 2AU. \citet{2017ApJ...842..118L} found that while their non-equilibrium models matched T$_{\rm eff}$ estimates from \citet{2015ApJ...804...92S}, the non-equilibrium models resulted in a significantly lower gravity for this object.\\
    \noindent
    \textbf{WISEA J140518.32+553421.3 (W1405, Y0.5):} W1405 is an early-Y dwarf that was identified by a methane-induced H-band feature \citep{2011ApJS..197...19K,2011ApJ...743...50C}. \citet{2012ApJ...756..172M} obtained ground-based YJH photometry in order to compare with a suite of models which incorporated various condensates including several sulfides, KCl, and Cr. They found that these models better fit near-infrared data better than completely cloud-free models. Using updated YJHK photometry, \citet{2013ApJ...763..130L} found that this object should be cool enough to display effects from the presence of water clouds, but their model grid did not extend down to cool enough temperatures for a reliable water-cloud model fit. \citet{2013A&A...550L...2L} obtained lower-limit z-band measurements for this dim object. \citet{2015ApJ...804...92S} reclassified this object as Y0.5 due to its J-band spectroscopy being narrower than the Y0 spectral standard. \citet{2016ApJ...823..152C} obtained Spitzer 3.6 and 4.5$\mu m$ light curves and found the first evidence for variability in a Y dwarf with 3.6\% variability detected with an 8.2hr period. \citet{2017ApJ...842..118L} found this object to also be consistent with solar metallicity and age. \\
    \noindent
    \textbf{WISE J154151.65-225024.9  (W1541, Y1):} W1541 is the latest-type Y dwarf we have analyzed which was re-classified as Y1 based upon the width of J-band spectroscopic measurements \citep{2015ApJ...804...92S}. This object is another Y dwarf part of the initial WISE discovery papers \citep{2011ApJS..197...19K, 2011ApJ...743...50C,2012ApJ...753..156K}. \citet{2012ApJ...756..172M} obtained ground-based Y and J-band photometry of this object where they note that their cloudy, rather than cloud-free models better reproduce the observed colors of their Y dwarf photometry. However, \citet{2012ApJ...750...74S} noted that with improved NH$_3$ opacities, their cloud-free models well-matched the observed colors. Ground-based YJHK photometry was obtained by \citet{2013ApJ...763..130L} who's cloud-free models estimate T$_{eff} \sim $325K. \citet{2015ApJ...799...37L} obtained improved H-band measurements and compared measured colors with a suite of both water-cloud and cloud-free models. Though they find that the inclusion of water-clouds do better-fit several colors, there are still magnitude-scale systematic offsets between the models and data. \citet{2017ApJ...842..118L} was able to successfully reproduce either Y or J-band spectroscopy but a simultaneous fit to the entire YJH spectrum could not be obtained.

\section{Results} \label{sec:results}

\begin{figure*}
        \centering
        \includegraphics[width=\textwidth]{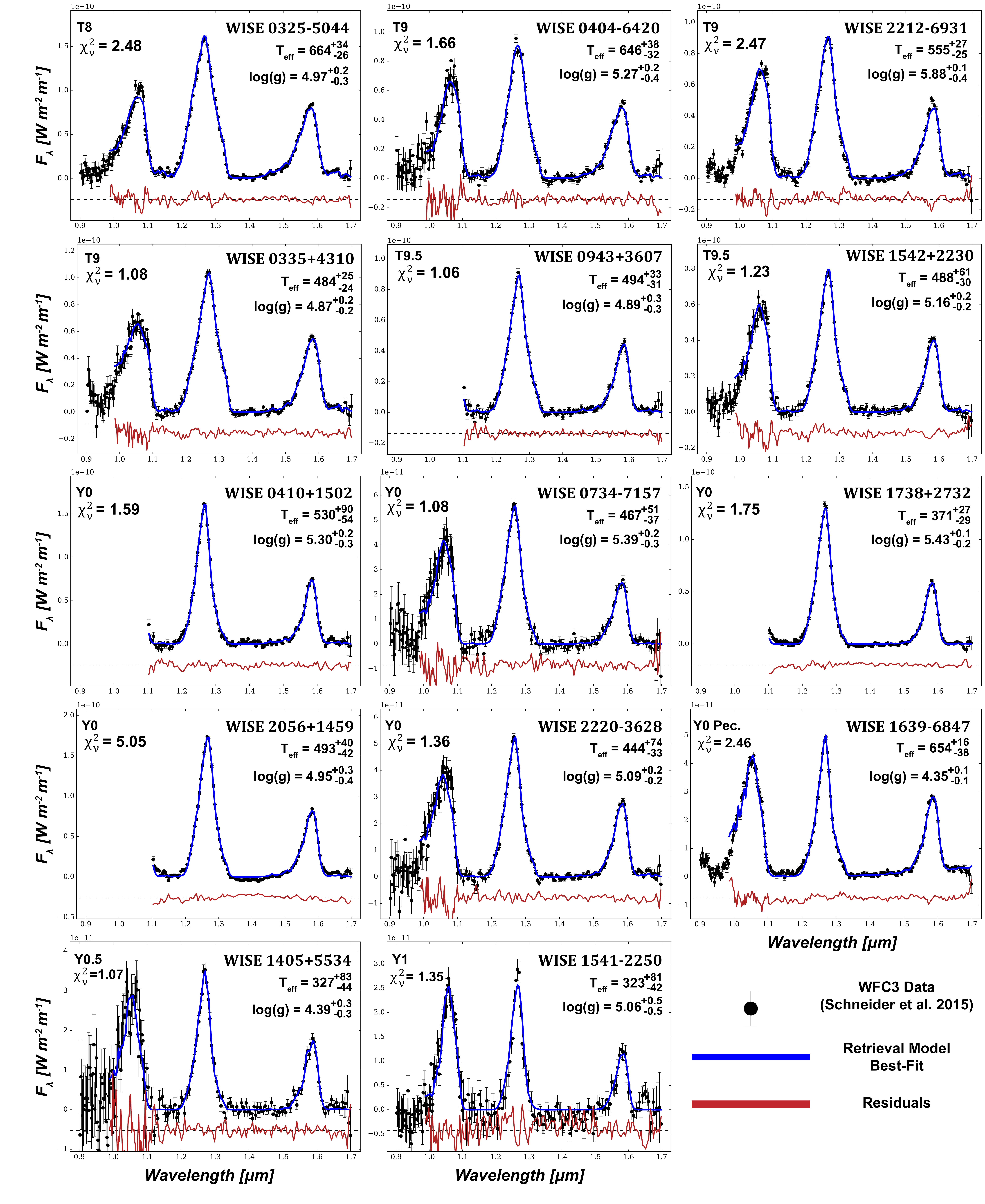}
        \caption{WFC3 observations (black points), best-fit retrieval model (blue), and residuals (lower, red) for the WFC3 sample sorted by the \citet{2015ApJ...804...92S} spectral classification (upper left). The retrieved log(g) [cgs] and derived effective temperature [K] are given in the upper right hand corner of each panel.}
        \label{fig:bigspec}
    \end{figure*}

Here we present our results from the analysis of our 14 late-T/early-Y dwarfs. Full posteriors of all model parameters are available online\footnote{Zenodo link TBD.}. Figure \ref{fig:bigspec} summarizes the fits with the WFC3 data in black, best-fit spectra in blue, and residuals in red. Several objects only have 1.1-1.7$\mu m$ coverage as full YJK coverage requires both G105 and G141 spectra from HST \citep{2015ApJ...804...92S}. From visual inspection, there is no systematic structure in the residuals and most of our objects have a $\chi^2_{\nu}$ between 1 and 3. W2056 has a higher $\chi^2_{\nu}$=5.05 but this is due to an oversubtraction artifact with fluxes between the J and H bands being well below 0 by $\sim2\sigma$. We have experimented with removing such data from our fit, but found it to not impact our results (see Section \ref{sec:gravity}).

We first discuss the implications of our constraints on the temperature structure and evolutionary properties of these objects. Our retrieved evolutionary parameters are enumerated in Table \ref{tab:evolution}. We then highlight our retrieved abundances which are listed in Table \ref{tab:abundance}. Finally we discuss how observations of these objects with JWST will impact our ability to characterize these objects.

\subsection{Vertical Temperature Structure} \label{sec:temp}

	 \begin{figure*}
        \centering
        \includegraphics[width=\textwidth]{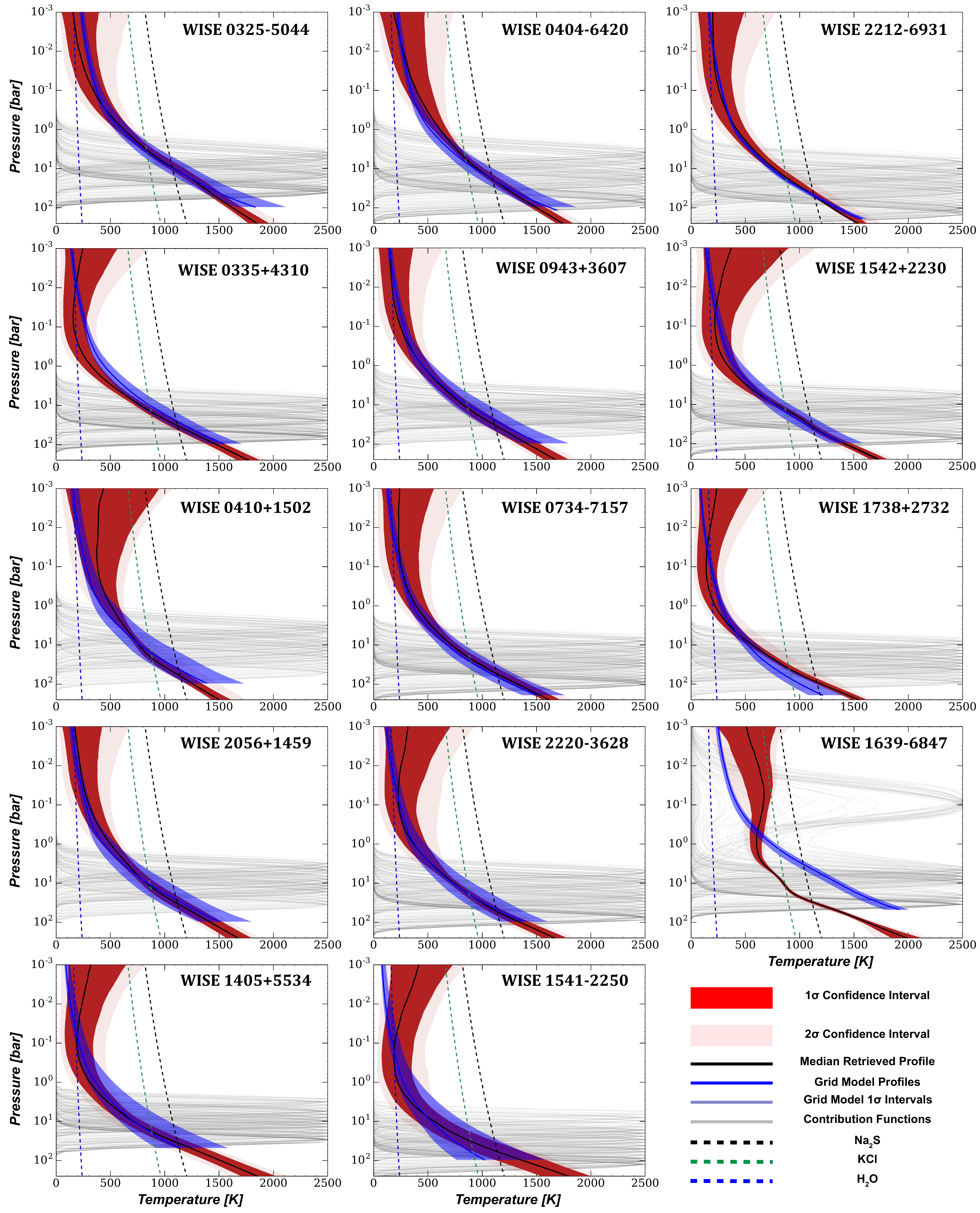}
        \caption{Retrieved TP profiles for all targets. Black lines are median values with red 1$\sigma$ and 2$\sigma$ confidence intervals. In grey are the contribution functions of the atmosphere from the WFC3 observations and can be thought of as the effective photosphere. Overlaid (blue) are solar-composition radiative-convective equilibrium profile spreads derived from the retrieved spread in effective temperature and gravity. Finally we also include solar-composition equilibrium condensation curves (dashed lines) for several important species \citep{2012ApJ...756..172M,2014ApJ...787...78M}. Most systems' retrieved TP profiles are in good agreement with radiative-convective equilibrium, save for W1639 (see text).}
        \label{fig:bigtp}
    \end{figure*}

	For any non-irradiated sub-stellar object, the energy balance, and hence thermal structure of the atmosphere, is governed by the flow of internal heat flux through the atmosphere, controlled by the gravity and atmospheric opacity. These properties are directly set by both the mass and age of the system \citep[e.g.][]{1996ApJ...465L.123A}. In ultra cool dwarfs, energy is primarily transported through radiation and convection \citep{2015ARA&A..53..279M}.
    
    Late-T/early-Y type objects are ideal for the characterization of the TP profile structure due to the high degree of spectral modulation (which maps to a wide range of probed pressures) and the presumed lack of optically thick clouds \citep{2012ApJ...756..172M}. As before, we make few a priori assumptions regarding the thermal structure of the atmosphere, and instead allow the observations to drive the solutions. By then making comparisons between our results, and those of self-consistent models, we can then investigate where ``atypical" atmospheric processes, such as deviations from radiative-convective equilibrium, may be occurring. Should any significant deviations be found, such information can be utilized in order to improve grid-based models' treatment of atmospheric structure by inclusion of other possible sources of heating. \citep[e.g.][]{2014MNRAS.440.3675S,2015ApJ...804L..17T}.
    
    Figure \ref{fig:bigtp} summarizes the resulting TP profiles. The median TP profile is shown in black, with 1$\sigma$ and 2$\sigma$ confidence intervals outlined in red. Overlaid on top of the retrieval results are radiative-convective equilibrium profiles (blue) derived using the ScCHIMERA modeling tool described in \citet{2018AJ....156..133P} and \citet{2018A&A...618A..63B}, generated using the T$_{\rm eff}$ and log(g) range derived from the retrieval results. We compute the effective temperature as in Parts I and II by numerically integrating for the bolometric flux of the retrieved spectral spread for each object from 1 - 20 $\mu$m. Contribution functions (grey) from the WFC3 observations are also overlaid which can be treated as the effective photosphere probed by WFC3. As was noted in Part II, we reiterate that the atmospheric structure above and below these regions is largely driven by our TP profile smoothing parameter (described in Part I), and that interpretation of such structure should be done with caution.
    
    The WFC3 observations probe pressures from roughly 1-100 bars with typical 1$\sigma$ temperature uncertainties of $\sim$50-100 K, consistent with the SpeX T-dwarfs from Part II.  For a large majority of objects, the retrieved structures appear consistent with the assumption of radiative-convective equilibrium. Though true for a large majority of objects, W1639 stands out in stark contrast. The retrieved TP profile shows almost no consistency with that of radiative-convective equilibrium, as well as an interesting ``kink" structure at roughly 10bars. The unique atmospheric structure also suggests our WFC3 wavelengths are sensitive to much lower pressures of roughly 0.1-0.01bars. 
    
    
    \subsubsection{WISEA J1639-6847}
    
    W1639 is the only object in our sample with a classification of ``Y0: Peculiar'' \citep{2015ApJ...804...92S}. Though the object's J-band spectra well-match the Y0 spectral standard, the overall Y-J color is significantly bluer due to a bright Y-band. Additionally, the overall position of the Y-band is significantly blue-shifted when compared to the T9 spectral standard. These features, remarked upon by \citet{2015ApJ...804...92S} are not just an artifact of the WFC3 data, as the bright Y-band has also been observed with the ground-based FIRE instrument \citep{2012ApJ...759...60T}. These unique properties motivated \citet{2015ApJ...804...92S} to invest more of their limited WFC3 time to this object, resulting in much better Y-band constraints than the other objects in our sample (see Figure \ref{fig:bigspec}) and are the main contributor for the improved TP profile constraints relative to our other objects. Though the W1639 spectrum seems well-fit by our model (reduced $\chi^2$ of $\sim$2) there are several lingering issues with the resulting best-fit parameters.
    
    First is that, despite the object being in the Y0 Pec. classification, we derive an effective temperature of 654$^{+16}_{-38}$K. There has been a suggestion that double-diffusive convection can result in much shallower thermal structures than predicted by equilibrium grid models \citep{2015ApJ...804L..17T}. This would be consistent with the profile we retrieve for W1639. However, using basic energy balance arguments and thermodynamics, \cite{2018ApJ...853L..30L} demonstrated that this mechanism would result in {\it steeper}, not shallower, temperature gradients, in contrast to \cite{2015ApJ...804L..17T} and the profile of W1639. Release of latent heat due to condensation of various cloud species may also result in shallower adiabats. However Figure \ref{fig:bigtp} shows that the main contributor to such heat, water, would negligibly impact our objects based on the intersection of the retrieved TP profile with water's equilibrium condensation curve. 
    
    Additionally, our estimate for log(g) is the lowest out of our 11 objects at log(g)=4.35$^{+0.1}_{-0.1}$ requiring a radius of R=0.4$^{+0.03}_{-0.02}$R$_{\rm Jup}$ and M=1.5$^{+0.3}_{-0.3}$M$_{\rm Jup}$ (see Section \ref{sec:gravity}, Table \ref{tab:evolution}). These constraints are significantly smaller than allowed by typical field dwarfs and, combined with the peculiar TP profile, leads us to conclude that our data-driven retrieval model may not be well-suited to explain the physical characteristics of this single unique object. However we reiterate that the remaining object's TP profiles seem to agree well with assumptions of radiative-convective equilibrium.


\subsection{Effective Temperature, Gravity, Radius, \& Mass} \label{sec:gravity}

	\begin{table*}
		\renewcommand{\thetable}{\arabic{table}}
		\centering
		\caption{Retrieved \& Derived Evolutionary Parameters.} \label{tab:evolution}
		\begin{tabular}{ccccccc}
		\tablewidth{0pt}
        \hline
		\hline
		\decimals
        WISE/ALLWISE Name & Spec. Type & T$_{\rm eff}$ [K] & log(g) [cgs] & $R$ [R$_{\rm Jup}$] & Mass [M$_{\rm Jup}$] & Priors\\
        \hline
        WISEA J032504.52-504403.0 & T8 & 664$^{+34}_{-26}$ & 4.97$^{+0.19}_{-0.30}$ & 1.08$^{+0.11}_{-0.11}$ & 44$^{+24}_{-23}$ & Free\\
        & & 660$^{+29}_{-24}$ & 5.06$^{+0.27}_{-0.36}$ & 1.10$^{+0.11}_{-0.10}$ & 56$^{+46}_{-32}$ & Constrained\\
        \\
		WISEA J040443.50-642030.0 & T9 & 646$^{+38}_{-32}$ & 5.27$^{+0.18}_{-0.30}$ & 0.78$^{+0.06}_{-0.06}$ & 47$^{+22}_{-24}$ & Free\\
		& & 639$^{+37}_{-30}$ & 5.20$^{+0.22}_{-0.43}$ & 0.81$^{+0.06}_{-0.06}$ & 42$^{+24}_{-25}$ & Constrained\\
		\\
		WISEA J221216.27-693121.6 & T9 & 555$^{+27}_{-25}$ & 5.88$^{+0.08}_{-0.35}$ & 0.47$^{+0.05}_{-0.03}$ & 69$^{+8}_{-34}$ & Free \\
		& & 540$^{+40}_{-32}$ & 5.25$^{+0.16}_{-0.29}$ & 0.71$^{+0.02}_{-0.02}$ & 36$^{+16}_{-17}$ & Constrained\\
		\\
		WISEA J033515.07+431044.7 & T9 & 484$^{+25}_{-24}$ & 4.87$^{+0.22}_{-0.22}$ & 0.87$^{+0.06}_{-0.06}$ & 23$^{+15}_{-10}$ & Free \\
		& & 483$^{+24}_{-25}$ & 4.87$^{+0.23}_{-0.21}$ & 0.88$^{+0.06}_{-0.06}$ & 23$^{+14}_{-9}$ & Constrained\\
		\\
        WISEA J094306.00+360723.3 & T9.5 & 494$^{+33}_{-31}$ & 4.89$^{+0.30}_{-0.31}$ & 0.70$^{+0.07}_{-0.07}$ & 15$^{+15}_{-8}$ & Free \\
		& & 494$^{+36}_{-36}$ & 4.86$^{+0.30}_{-0.30}$ & 0.75$^{+0.07}_{-0.06}$ & 16$^{+17}_{-8}$ & Constrained\\
		\\
        WISEA J154214.00+223005.2 & T9.5 & 488$^{+60}_{-30}$ & 5.16$^{+0.15}_{-0.18}$ & 0.61$^{+0.05}_{-0.05}$ & 21$^{+11}_{-8}$ & Free \\
		& & 484$^{+39}_{-26}$ & 5.07$^{+0.16}_{-0.26}$ & 0.71$^{+0.05}_{-0.04}$ & 23$^{+12}_{-10}$ & Constrained\\
		\\
		WISEA J041022.75+150247.9 & Y0 & 530$^{+90}_{-54}$ & 5.30$^{+0.23}_{-0.32}$ & 0.73$^{+0.10}_{-0.08}$ & 43$^{+24}_{-21}$ & Free\\
		& & 529$^{+83}_{-86}$ & 5.06$^{+0.29}_{-0.59}$ & 0.75$^{+0.07}_{-0.04}$ & 27$^{+24}_{-13}$ & Constrained\\
		\\
		WISEA J073444.03-715743.8 & Y0 & 467$^{+51}_{-37}$ & 5.39$^{+0.17}_{-0.28}$ & 0.71$^{+0.07}_{-0.07}$ & 50$^{+21}_{-23}$ & Free\\
		& & 456$^{+50}_{-34}$ & 5.24$^{+0.18}_{-0.33}$ & 0.77$^{+0.77}_{-0.05}$ & 42$^{+22}_{-22}$ & Constrained\\
		\\
		WISEA J173835.52+273258.8 & Y0 & 371$^{+27}_{-29}$ & 5.43$^{+0.13}_{-0.17}$ & 0.71$^{+0.05}_{-0.05}$ & 59$^{+15}_{-22}$ & Free\\
		& & 371$^{+33}_{-30}$ & 5.20$^{+0.2}_{-0.29}$ & 0.73$^{+0.04}_{-0.03}$ & 34$^{+20}_{-17}$ & Constrained\\
		\\
		WISEA J205628.88+145953.6 & Y0 & 493$^{+40}_{-42}$ & 4.95$^{+0.31}_{-0.35}$ & 0.67$^{+0.07}_{-0.05}$ & 16$^{+15}_{-8}$ & Free\\
		& & 485$^{+38}_{-38}$ & 4.93$^{+0.25}_{-0.29}$ & 0.72$^{+0.03}_{-0.02}$ & 18$^{+14}_{-9}$ & Constrained\\
		\\
		WISEA J222055.34-362817.5 & Y0 & 444$^{+74}_{-33}$ & 5.09$^{+0.24}_{-0.23}$ & 0.72$^{+0.08}_{-0.07}$ & 26$^{+21}_{-9}$ & Free\\
		& & 449$^{+57}_{-35}$ & 5.07$^{+0.16}_{-0.26}$ & 0.74$^{+0.07}_{-0.07}$ & 26$^{+13}_{-12}$ & Constrained\\
		\\
		WISEA J163940.84-684739.4 & Y0 Pec. & 654$^{+16}_{-38}$ & 4.35$^{+0.09}_{-0.08}$ & 0.40$^{+0.03}_{-0.02}$ & 1.5$^{+0.3}_{-0.3}$ & Free\\
		& & - & - & - & - & Constrained\footnote{\label{1639const}Retrieval model could not converge upon a physically realistic TP profile}\\
		\\
		WISEA J140518.32+553421.3 & Y0.5 & 327$^{+83}_{-44}$ & 4.39$^{+0.28}_{-0.31}$ & 0.66$^{+0.12}_{-0.09}$ & 4.4$^{+3.9}_{-2.1}$ & Free\\
		& & 338$^{+98}_{-58}$ & 3.89$^{+0.22}_{-0.15}$ & 0.75$^{+0.08}_{-0.06}$ & 1.7$^{+1.4}_{-0.5}$ & Constrained\\
		\\
		WISE J154151.65-225024.9 & Y1 & 323$^{+81}_{-42}$ & 5.06$^{+0.50}_{-0.48}$ & 0.33$^{+0.07}_{-0.05}$ & 5.4$^{+9.7}_{-3.4}$ & Free\\
		& & 389$^{+85}_{-87}$ & 3.91$^{+0.33}_{-0.19}$ & 0.72$^{+0.04}_{-0.02}$ & 1.7$^{+2.0}_{-0.6}$ & Constrained\\
		\hline
	\end{tabular}
	\end{table*}

    \begin{figure}
	    \centering
         \includegraphics[width=\linewidth]{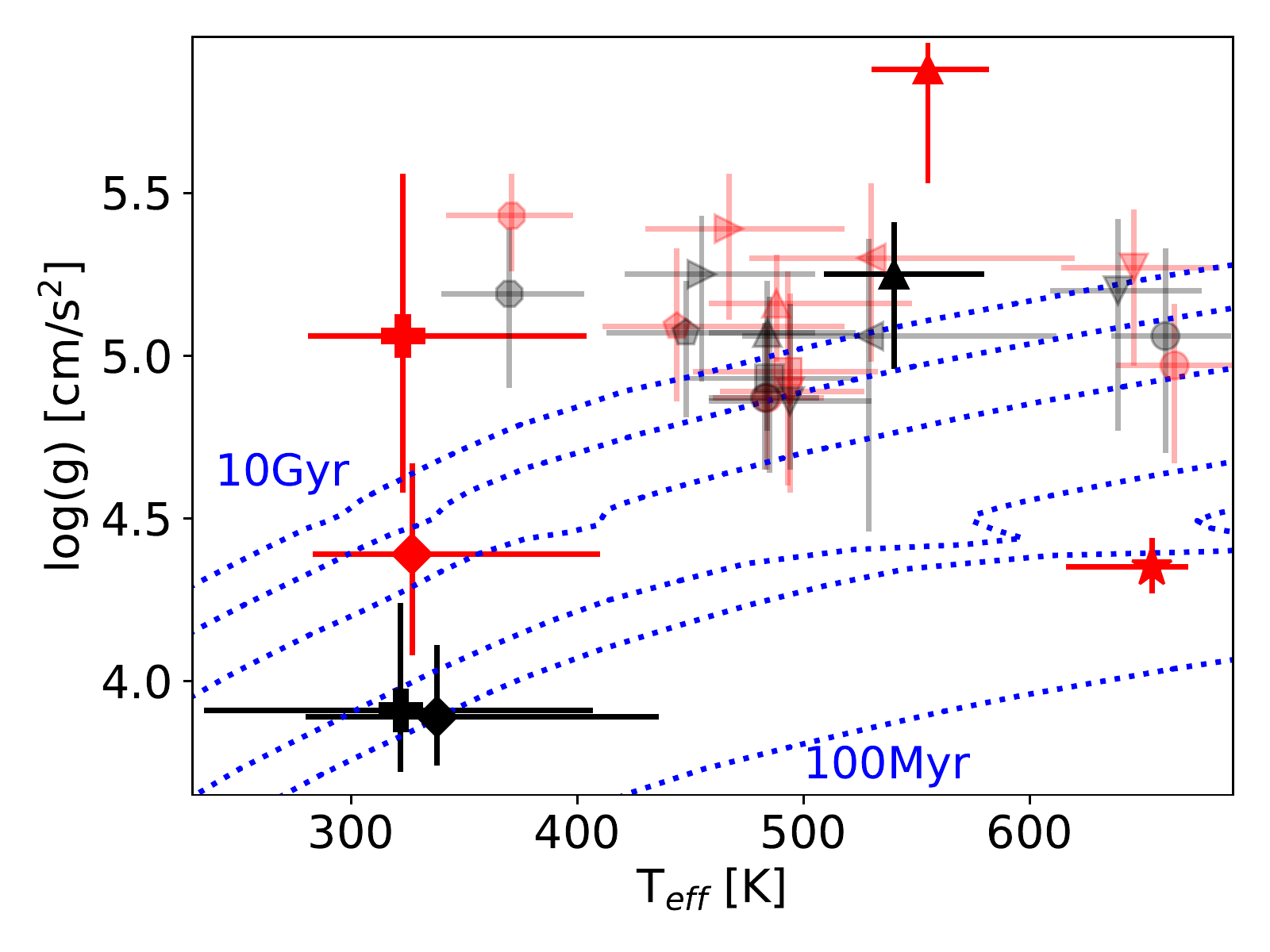}
         \caption{T$_{\rm eff}$ and log(g) 1$\sigma$ constraints for our free retrieval results (red) and constrained retrieval (black). Each object has its own unique symbol. Most objects, save the coldest two (W1405,W1541) and W1639 (see text) show consistent results between the free and constrained retrievals at 1$\sigma$. Evolutionary trends from Marley et al. (2019, in prep.) with constant age (blue dotted) are overlaid for context at 100Myr, 600Myr, 1Gyr, 3Gyr, 6Gyr, and 10Gyr.}
         \label{fig:logg_teff}
    \end{figure}

The effective temperature, gravity, and thus radius, and mass are diagnostic of brown dwarf evolutionary history \citep[e.g.][]{2001RvMP...73..719B, 2003A&A...402..701B, 2008ApJ...689.1327S}. Evolution models suggest that our late-T ($\geq$T8) and early-Y ($\leq$Y1) sample, should have T$_{\rm eff}$'s from 800-350K \citep{2013ApJS..208....9P}.  Field-age late-T and early-Y dwarfs are expected to have log(g)$\approx$5 with a relatively strong upper bound at log(g)$\approx$5.3 for even the oldest and coldest brown dwarfs possible \citep[e.g][]{2008ApJ...689.1327S}.  Field-aged objects over the 10 - 80 M$_{\rm Jup}$ range are expected to have radii within $\sim$20\% of Jupiter's. 

In this work, the total ($R/D$)$^2$ scaling factor is a free parameter. By using constraints on the distance $D$ from the literature and our retrieved constraints on ($R/D$)$^2$, we are then able to \textit{derive} constraints on the photometric radius $R$. Using this \textit{derived} constraint, with our retrieved log(g), we can then derive the total mass of the object $M$, with a prior upper limit of 80M$_{\rm Jup}$. Table \ref{tab:evolution} and Figure \ref{fig:logg_teff} summarize the retrieved and derived estimates for these evolutionary parameters under two sets of model assumptions.


\subsubsection{Free Retrieval}

We first focus on our a less constrained, ``free'' retrieval, which only incorporates the 80M$_{ \rm Jup}$ mass prior upper limit, as has been done in Parts I and II. Our retrieved log(g) and derived T$_{\rm eff}$ for this case are shown as red symbols in Figure \ref{fig:logg_teff}. In general we find that the uncertainties in both log(g) and T$_{\rm eff}$ are consistent with the results from Part II with 1$\sigma$ errors between 0.1-0.5 dex and 30-90K respectively. This is encouraging as both our dataset in this work, and the dataset in Part II had comparable SNR on the observed spectra. For a majority of objects, our derived effective temperatures agree with the spectral types given in \citet{2015ApJ...804...92S} when compared to the table provided in \citet{2013ApJS..208....9P}. The notable exception to this trend is W1639 whose unique TP profile is discussed in Section \ref{sec:temp}.

\begin{figure}
        \centering
        \includegraphics[width=1.05\linewidth]{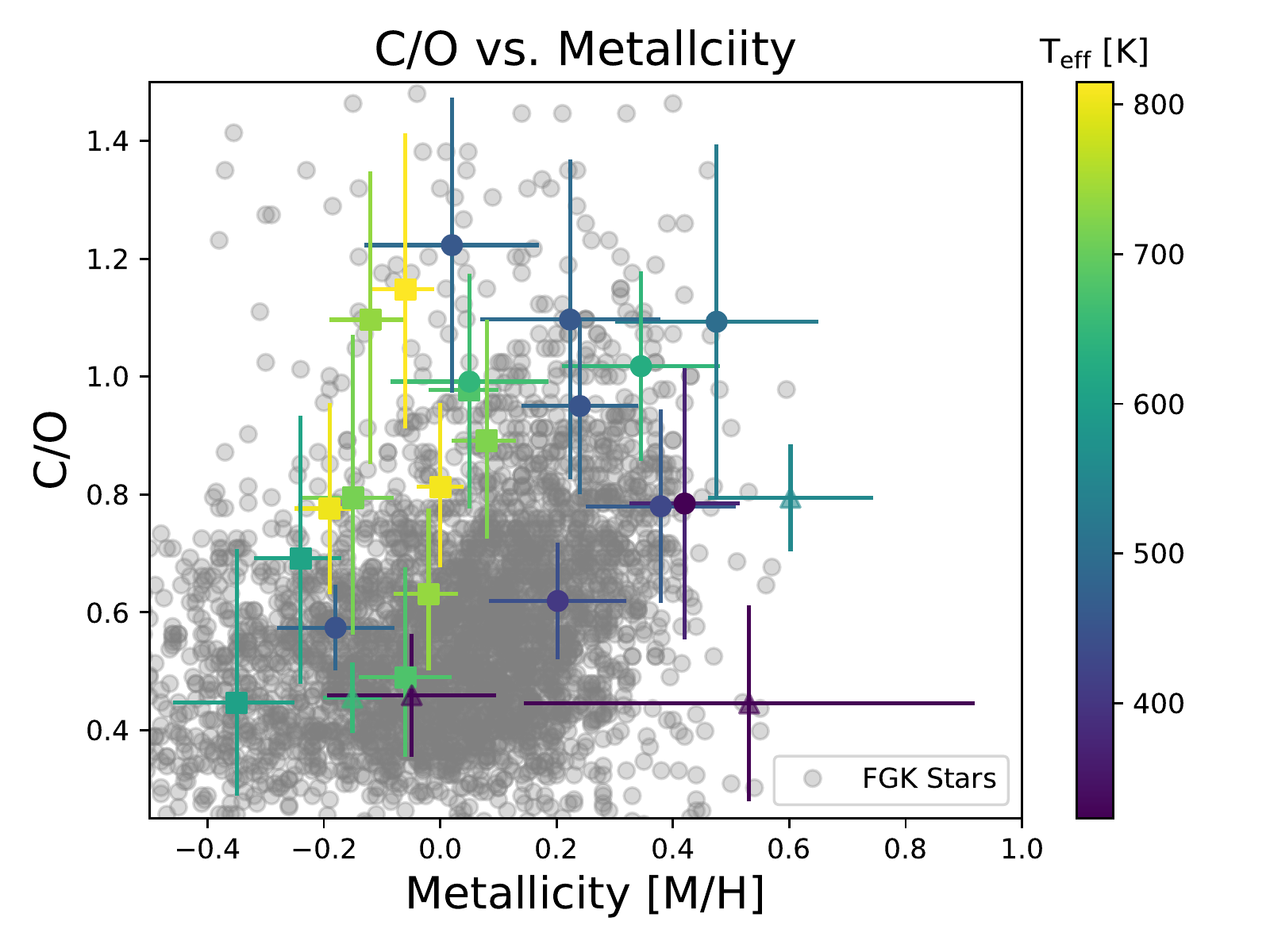}
        \caption{Metallicity vs. C/O for our retrieved results (circles) and the results from Part II (squares). Triangles should be interpreted with caution (see text). We also overlay estimates for the local FGK stellar population (grey) for context \citep{2014AJ....148...54H}. Our combined sample of late-T and early-Y dwarfs seem relatively consistent with the local stellar population.}
        \label{fig:MetCO}
	\end{figure}

Overplotted on Figure \ref{fig:logg_teff} are curves of constant age (blue,dashed) from the upcoming Sonora grid of evolutionary models (Marley et al. 2019, in prep.). Several objects appear to have gravities that extend well beyond those anticipated by the 10Gyr curve, though our retrieved uncertainties are large enough to be consistent with the oldest models at $\sim$2$\sigma$. This result is not unique to our retrieval approach as \citet{2015ApJ...804...92S} also noted that grid-model comparisons resulted in a similar result with several object's log(g) estimates requiring gravity values higher than computed in their grid model. In addition to high gravity estimates, we also found several objects to have smaller radii than expected for our late-T, early-Y sample with several objects below a lower limit value of $\sim$0.7R$_{\rm Jup}$ \citep{2008ApJ...689.1327S}. Both the uncomfortably high gravities and small radii prompted us to explore the robustness of our results through a battery of tests. 

We first investigated the possibility of an unknown systematic error in the observed spectra biasing the fits. We considered this as it was noticed that the observed fluxes fell below 0, well outside of the reported 1-sigma uncertainties in some cases, in the deep absorption bands for several objects (namely W2056, W2220, W1541, see Figure \ref{fig:bigspec}). This is due to oversubtraction when attempting to remove background fluctuations. To test this we introduced a free parameter to uniformly shift the model spectra fluxes to within the 1$\sigma$ error estimates of the data, but found that it did not produce any considerable change in parameter estimates.

Next we explored some of the key assumptions made within our model. In this initial set of ``free'' retrievals we had assumed a hard prior upper limit of 80M$_{\rm Jup}$ on the mass. This prior rejects combinations of radii and gravities that would exceed this mass limit. We tested the sensitivity of our results to this mass upper limit by effectively removing the prior. These resulted in significantly higher retrieved gravities with some objects such as W1738 reaching as high as log(g)=5.7$^{+0.18}_{-0.32}$. This indicated to us that the data was indeed favoring higher masses and gravities along with lower radii.

We then explored how radius assumptions (through $(R/D)^2$--assuming a distance) could influence the retrievals by fixing the radii to a realistic 0.9R$_{\rm Jup}$ (and turning the mass upper limit prior back on).  For the case of W0734 (originally retrieved R$\sim$0.71R$_{\rm Jup}$, log(g)=5.39$^{+0.17}_{-0.28}$), we found that the fixed radius resulted in a decreased gravity (log(g)=5.12$^{+0.2}_{-0.33}$). However, the resulting marginalized posterior is highly non-Gaussian due to the enforcement of the 80M$_{\rm Jup}$ cutoff. This suggests that, despite enforcing these priors, the high-gravity solution is still favored.


\subsubsection{Constrained Retrieval}

We finally decided to run a completely separate set of retrievals on all of our objects with more stringent priors based on results from evolutionary models that we have labeled as ``constrained'' retrievals. This included restricting 0.7R$_{\rm Jup}<$R$<$2.0R$_{\rm Jup}$, 3.5$<$log(g)$<$5.5, and effectively removing the mass upper limit. The results of this analysis are shown as black symbols in Figure \ref{fig:logg_teff} and are enumerated in Table \ref{tab:evolution}. Objects whose ``constrained" retrieval results are within 1$\sigma$ of the ``free" retrieval are translucent, while those who log(g) change by $>1\sigma$ are opaque.

We obtained two key results from this constrained retrieval test. First is that, regardless of our priors on evolutionary parameters, the data suggest that these objects have anomalously high gravity estimates. This can be seen in Figure \ref{fig:logg_teff} where most objects still lie above the 10Gyr trend for both the ``constrained" (black) and ``free" (red) retrieval. For the Y0 objects, we find a consistent decrease in their log(g) estimates by upwards of 0.3 dex and a slight increase in the radii estimates to roughly 0.75R$_{\rm Jup}$. Though this places our retrieval results in better agreement with evolutionary models, our posterior distributions for log(g) are consistently non-Gaussian, and push against the log(g)=5.5 upper limit, suggesting that the high gravity solution is still favored.

\begin{table*}
		\renewcommand{\thetable}{\arabic{table}}
		\centering
		\caption{Retrieved Atmospheric Abundances} \label{tab:abundance}
		\begin{tabular}{cccccccccc}
		\tablewidth{0pt}
        \hline
		\hline
		\decimals
        WISE/ALLWISE Name & Spec.& H$_2$O\footnote{\label{vmr}All abundances are reported as the log of the volume mixing ratio log(VMR) where the remainder of the gas is taken to be H$_2$-He at a fixed solar ratio.} & CH$_4$\footref{vmr} &CO\footref{vmr}\footnote{\label{upperlim}All of these measurements represent 3$\sigma$ upper limits (see text).} & CO$_2$\footref{vmr}\footref{upperlim} & C/O\footnote{These are not relative to solar. Solar [C/O] is 0.55 in this table.} & H$_2$S\footref{vmr}\footref{upperlim} & NH$_3$\footref{vmr} & Na+K\footref{vmr} \\
        &Type&&&&&&&&\\
        \hline
        WISEA J032504.52-504403.0 & T8 & -3.31$^{+0.12}_{-0.13}$ & -3.05$^{+0.11}_{-0.16}$ & -4.1 & -3.7 & 0.99$\pm0.18$ & -5.0 & -4.49$^{+0.12}_{-0.18}$ & -5.52$^{+0.09}_{-0.07}$ \\
		WISEA J040443.50-642030.0 & T9 & -3.01$^{+0.11}_{-0.13}$ & -2.74$^{+0.10}_{-0.16}$ & -3.0 & -3.3 & 1.02$\pm$0.16 & -5.0 & -4.63$^{+0.13}_{-0.20}$ & -6.0\footref{upperlim} \\
		WISEA J221216.27-693121.6 & T9 & -2.59$^{+0.07}_{-0.18}$ & -2.56$^{+0.05}_{-0.16}$ & -2.9 & -3.3 & 0.79$\pm$0.09 & -6.8 & -4.05$^{+0.08}_{-0.13}$ & -5.0\footref{upperlim} \\
		WISEA J033515.07+431044.7 & T9 & -3.35$^{+0.09}_{-0.09}$ & -3.48$^{+0.11}_{-0.11}$ & -3.8 & -3.9 & 0.57$\pm$0.07 & -5.3 & -4.78$^{+0.13}_{-0.12}$ & -5.97$^{+0.07}_{-0.10}$ \\
        WISEA J094306.00+360723.3 & T9.5 & -3.35$^{+0.14}_{-0.15}$ & -3.13$^{+0.17}_{-0.15}$ & -3.3 & -3.2 & 1.22$\pm$0.25 & -5.0 & -4.46$^{+0.16}_{-0.16}$ & -5.2\footref{upperlim} \\
        WISEA J154214.00+223005.2 & T9.5 & -3.04$^{+0.09}_{-0.08}$ & -2.92$^{+0.08}_{-0.10}$ & -4.2 & -4.3 & 0.95$\pm$0.15 & -6.0 & -4.32$^{+0.10}_{-0.12}$ & -6.7\footref{upperlim} \\
		WISEA J041022.75+150247.9 & Y0 & -2.90$^{+0.13}_{-0.15}$ & -2.63$^{+0.17}_{-0.19}$ & -3.3 & -4.1 & 1.09$\pm$0.30 & -4.3 & -4.11$^{+0.15}_{-0.19}$ & -5.0\footref{upperlim} \\
		WISEA J073444.03-715743.8 & Y0 & -2.91$^{+0.12}_{-0.15}$ & -2.77$^{+0.09}_{-0.14}$ & -3.4 & -3.7 & 0.78$\pm$0.16 & -6.0 & -4.29$^{+0.10}_{-0.14}$ & -6.0\footref{upperlim} \\
		WISEA J173835.52+273258.8 & Y0 & -2.87$^{+0.08}_{-0.08}$ & -2.75$^{+0.12}_{-0.10}$ & -3.3 & -4.1 & 0.79$\pm$0.23 & -5.0 & -4.21$^{+0.10}_{-0.09}$ & -5.2\footref{upperlim} \\
		WISEA J205628.88+145953.6 & Y0 & -3.18$^{+0.16}_{-0.15}$ & -2.89$^{+0.18}_{-0.17}$ & -4.2 & -4.4 & 1.10$\pm$0.27 & -5.0 & -4.44$^{+0.17}_{-0.17}$ & -5.5\footref{upperlim} \\
		WISEA J222055.34-362817.5 & Y0 & -3.04$^{+0.11}_{-0.10}$ & -3.00$^{+0.11}_{-0.12}$ & -4.2 & -4.3 & 0.62$\pm$0.10 & -5.8 & -4.19$^{+0.08}_{-0.10}$ & -6.8\footref{upperlim} \\
		WISEA J163940.84-684739.4 & Y0Pec. & -3.32$^{+0.04}_{-0.04}$ & -3.42$^{+0.05}_{-0.04}$ & -4.3 & -4.6 & 0.46$\pm$0.06 & -6.3 & -4.72$^{+0.05}_{-0.04}$ & -7.0\footref{upperlim} \\
		WISEA J140518.32+553421.3 & Y0.5 & -3.24$^{+0.15}_{-0.13}$ & -3.33$^{+0.14}_{-0.16}$ & -3.6 & -3.8 & 0.46$\pm$0.10 & -5.0 & -4.84$^{+0.14}_{-0.16}$ & -6.0\footref{upperlim} \\
		WISE J154151.65-225024.9 & Y1 & -2.68$^{+0.26}_{-0.24}$ & -2.80$^{+0.26}_{-0.26}$ & -3.5 & -3.6 & 0.45$\pm$0.17 & -5.0 & -4.43$^{+0.21}_{-0.23}$ & -6.4\footref{upperlim} \\
		\hline
	\end{tabular}
	\end{table*}

Our second result is that, regardless of our priors on evolutionary parameters, we still obtain the same constraints on our chemical abundances to within 1$\sigma$. This was a bit surprising as there is a well-known correlation between the gravity and overall metallicity for these objects. We ensured this by picking three of our objects with anomalously high gravities (W2212, W0734, W1738) and enforcing that log(g)=5.0. We found that our retrieved metallicity did indeed decrease as expected, however our overall fit to the data was much worse under this assumption, with an average delta $\chi^2_{\nu}$ of 6.5, indicating that our original retrieved metallicities and high gravities are the statistically favored solution.


\subsubsection{Caveats \& Exceptions}

There were four objects in total which did not follow these trends which are the opaque points in Figure \ref{fig:logg_teff}. Though W2212 obtains plausible constraints on the mass and T$_{\rm eff}$, it requires a radius of R=$0.47^{+0.05}_{-0.03}$ under the ``free" retrieval assumption. Our ``constrained" retrieval does result in a more physically realistic R=$0.71^{+0.02}_{-0.02}$, we find that our constraints on the chemical abundances change by $\sim2\sigma$. We ran a separate retrieval on this target using an different distance estimate from \citet{2012ApJ...753..156K} where we obtain a physically realistic R=$0.68^{+0.06}_{-0.05}$ and our chemical abundances did not change beyond 1$\sigma$ though the retrieved gravity is still the largest of our sample at log(g)=$5.5^{+0.11}_{-0.17}$. Full model posteriors for this additional run are available at the linked Zenodo site.

For W1639, our retrieval model could not converge upon a physically realistic TP profile given the assumptions in the constrained retrieval and thus has no corresponding black star in Figure \ref{fig:logg_teff}. For both W1405 and W1541 (diamond and cross respectively) we find that though our retrieval model converges upon solutions for both objects, they are largely nonphysical. By enforcing stronger constraints on the radius, we find that a Jupiter-like mass and significantly lower gravities (by $\sim2\sigma$) are needed in order to well-match the spectra under these assumptions. Additionally, our constraints on the chemical abundances change by upwards of 3$\sigma$. Though we include these four objects in the results of subsequent sections, we strongly caution against over-interpretation of their chemical abundance constraints given they significantly change under different model assumptions.

The one technique which proved successful in reducing the retrieved gravity of an object without encountering non-Gaussian posteriors or changes in chemical abundances was changing the assumed parallax. Our distance estimates had been taken from two specific sources in the literature \citep{2018ApJ...867..109M,2019ApJS..240...19K}. These were chosen in order to use the most updated parallax estimates from the Spitzer instrument. However, several other campaigns have previously measured parallaxes for several of our targets \citep[e.g.][]{2016AJ....152...78L, 2017MNRAS.468.3764S}. In most cases, the distances proved consistent to our previous assumptions and, as expected, our retrieved parameters remained the same. However using the parallax measurement for W2056 from \citet{2017MNRAS.468.3764S} resulted in a more physically realistic log(g)=$4.58^{+0.33}_{-0.37}$. Though we did not find similar results for the other distance estimates from \citet{2017MNRAS.468.3764S}, we note that this result shows how sensitive our evolutionary parameters are to measured parallaxes. If the distance estimates are systematically biased in a similar fashion this may also account for the fact that our radii estimates are slightly lower than expected from evolutionary models.



\subsection{Composition} \label{sec:comp}
One of the key utilities of retrievals is their ability to directly determine the molecular abundances in an atmosphere, rather than assume them from elemental abundances and equilibrium chemistry. From the molecular abundances we can \textit{derive} the atomic abundance ratios (e.g., metallicity, C/O, N/O etc.), and more importantly, explore trends in these abundances which are diagnostic of atmospheric chemical mechanisms. The primary motivations for looking at molecular abundances in the Y-dwarf regime are to (1) determine at what temperature the alkali metals completely disappear and if it is consistent with grid-model chemical predictions, and to (2) determine the role of ammonia as it is anticipated to be strongly influenced by disequilibrium vertical mixing. Again, our retrieval forward model assumes constant-with-altitude (pressure) molecular mixing ratios. The retrieved abundances are therefore representative of column integrated abundances over the photosphere probed by WFC3.

\begin{figure*}
        \centering
        \includegraphics[width=\textwidth]{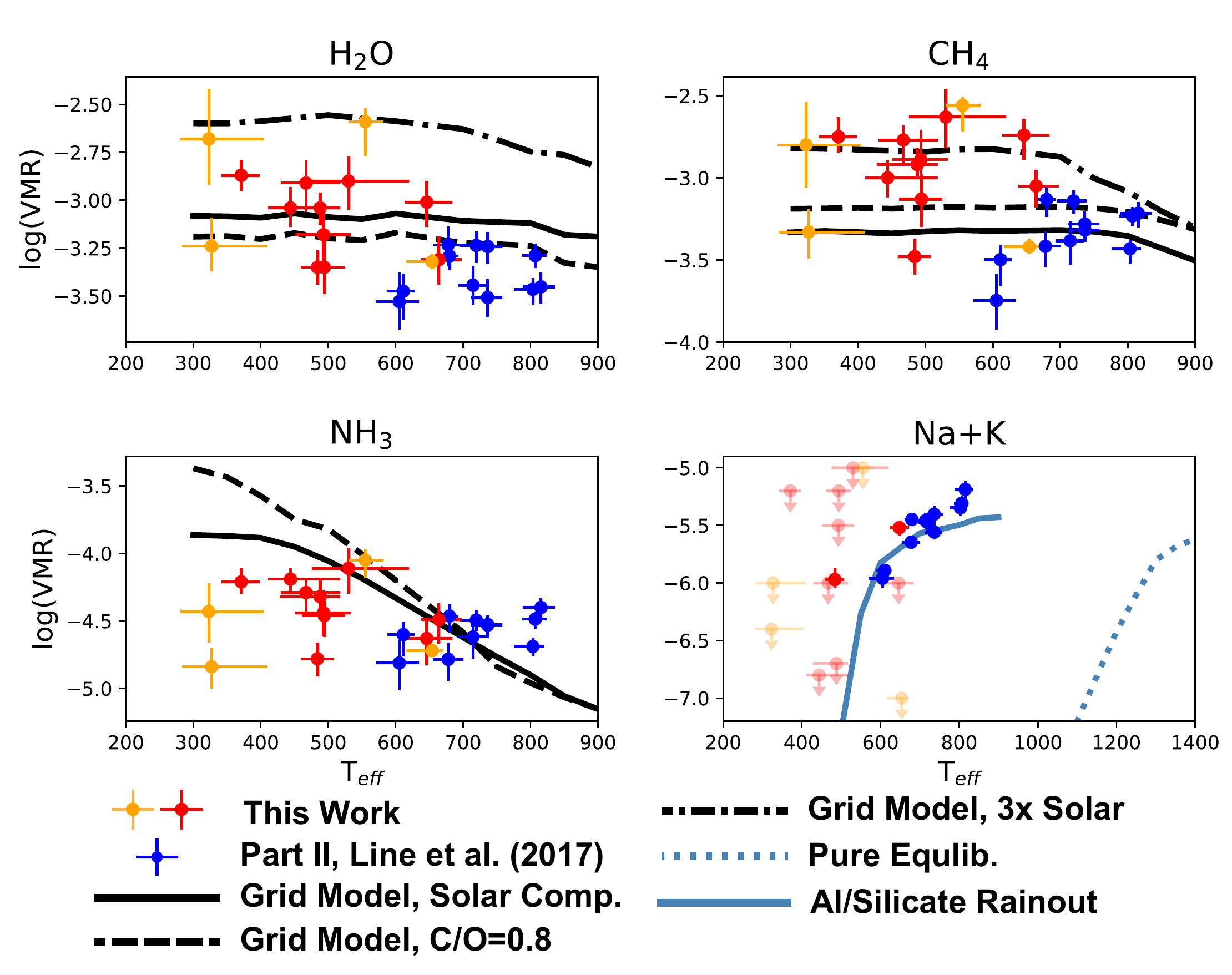}
        \caption{Constraints on our retrieved molecular abundances for H$_2$O (upper left), CH$_4$ (upper right), NH$_3$ (lower left) and Na+K (lower right) in units of Volume Mixing Ratio (VMR). Blue points are results from the hotter late-T sample in Part II. Red points are objects of this study whose abundances do not strongly depend on our assumptions of evolutionary priors (log(g), radius, mass) while yellow points are objects whose abundances are sensitive to these assumptions and should be interpreted with caution (see Section \ref{sec:gravity}). Overlaid are grid model profiles for various metallicites, C/O ratios, and rainout assumptions. Unless stated otherwise, curves are 1x solar composition with assumed thermochemical equilibrium. Pure equilibrium trend from \citet{2001RvMP...73..719B}.}
        \label{fig:bigabund}
    \end{figure*}

Table \ref{tab:abundance} summarizes the molecular abundance constraints (median and 68\% confidence interval). We find well defined, bounded constraints for H$_2$O, CH$_4$, NH$_3$, and in two cases Na+K, but obtain only upper limits for CO, CO$_2$, H$_2$S, and the alkalies. Upper limits are consistent with a non-detection as shown in Part II. These results are also broadly consistent with expectations from chemical equilibrium predictions as H$_2$O,  CH$_4$, and NH$_3$ are expected to be the dominant species where as CO, CO$_2$, H$_2$S, and the alkali metals less so \citep{1999ApJ...512..843B}. Section \ref{sec:chemtrends} highlights trends identified in both NH$_3$ and alkali metals.  First, we discuss the derived bulk atmospheric metallicity and carbon-to-oxygen ratios.


\subsubsection{Metallicity \& C/O}

The elemental abundance inventory of a substellar object is important to its evolutionary history as it governs total atmospheric opacity, and hence its cooling rate \citep{2001RvMP...73..719B}. It is important to understand the elemental abundances in brown dwarfs in order to place them into compositional context with both higher mass stars and lower mass planets.

One would expect the population of field brown dwarfs to have a similar elemental abundance pattern as stars, since both objects are thought to form via fragmentation within a molecular cloud. To contrast this, planets which are formed in protoplanetary disks can undergo migration within that disk. The existence of ice lines and dynamical models of migration have led to a range of predictions regarding planet-mass atmospheric elemental abundances. These can range any where from ``stellar composition'' to high metallicity ($>$100$\times$Solar, \citep[e.g.][]{2013ApJ...775...80F,2016ApJ...832...41M}) or high carbon-to-oxygen ratios (C/O $>$ 1, \citep[e.g.][]{2011ApJ...743L..16O,2014ApJ...794L..12M,2014Life....4..142H,2016A&A...595A..83E}). Identifying at what mass, in general, the diversity in composition substantially increases can ultimately assist us in truly bridging the gap between stars and planets. Since brown dwarfs sit between these mass limits, determining elemental abundances for a large number of substellar objects can help us in bridging this gap.

There are several challenging aspects to brown dwarf elemental abundance determinations. Firstly, at these cooler temperatures the chemical inventory is largely in the form of molecular, rather than elemental species. Molecules have much more complex spectroscopic features than atomic species with broad and deep roto-vibrational bands that overwhelm the spectral continuum; an oft used handle to obtain basic bulk parameters for hotter stars \citep[e.g.][]{2006ApJ...652.1604B}. Additionally, some molecular species are thought to be affected by both equilibrium condensate rainout and vertical disequilibrium mixing, while others can retain uniform chemical abundance profiles throughout the atmosphere \citep[e.g.][]{2001RvMP...73..719B,2007ApJS..168..140S}. Therefore, in order to accurately characterize the atmospheres of brown dwarfs, one must include the key molecular components covered over their bandpass, as well as the relevant chemical and dynamical processes which could affect such constituents.

Here, we focus our elemental abundance results on the metallicity and carbon-to-oxygen ratios only, as these are the most readily determinable elemental ratios for objects at these temperatures. Water and methane contain a bulk of the atmospheric metal content (C and O) for atmospheres cooler than $\le$1000K.

We determine directly from the retrieved molecular abundances the metallicity and C/O. The metallicity is computed by summing the molecular metal content (e.g., M=H$_2$O+2CO+3CO$_2$+NH$_3$+Na+K+CH$_4$+H$_2$S), then dividing by the background hydrogen content (H=2H$_2$+4CH$_4$+3NH$_3$+2H$_2$O+2H$_2$S) and finally normalizing by the solar M/H fraction to obtain our final ``metallicity" ([M/H] = log((M/H)/(M/H)$_{solar}$)). The C/O is determined dividing the total carbon (CO+CO$_2$+CH$_4$) by the total oxygen (H$_2$O+CO+2CO$_2$).

For both the metallicity and C/O, really, it is the water and methane that dominate.  We point out, as in Parts I and II, that this is a measure of the {\it atmospheric} elemental abundance inventory.  The bulk abundances can only be determined via chemical assumptions. Specifically, it is predicted that condensate rain out by silicates (enstatite, forsterite) can sequester oxygen by effectively locking it into condensates which "rain" out of the atmosphere and no longer react with the surrounding gas \citep[e.g.][]{1994Icar..110..117F}.  As in Part I we apply a correction factor to the C/O and metallicity by weighting the water abundance by a factor of 1.3 to accommodate for the lost O.

Figure \ref{fig:MetCO} shows our results of our retrieved metallicity and C/O constraints (circles) compared to the results for late-T dwarfs in Part II (squares), as well as a representative sample of these parameters from near-by FGK stars (grey circles) \citep{2014AJ....148...54H}. Overplotted (triangles) are the results for W2212, W1639, W1405, and W1541 for which the retrieved abundances, and thus C/O and metallicities, are dependent upon the choice of priors for evolutionary parameters and should be interpreted with caution (Section \ref{sec:gravity}). Plotted here are the results under the ``free" retrieval assumption to be consistent with the objects in Part II. 

We find that our metallicities are slightly enhanced, but overall broadly consistent with the local FGK stellar population and our C/O values are consistent with the results of Part II and the stellar population. We note that there appears to be no correlation between the effective temperature and metallicity or C/O for our sample. In Part II it was discussed that the apparent trend of increasing C/O with increasing metallicity for the late-T's could potentially be explained with super-solar [Si/O] ratios affecting the efficiency of oxygen rainout into silicates. By including our new late-T and early-Y sample, we find no such trend even if one were to discount the objects with questionable constraints.


\begin{table}
\renewcommand{\thetable}{\arabic{table}}
\centering
\caption{ScCHIMERA Grid Model Ranges and Step Sizes} \label{tab:gridmod}
\begin{tabular}{ccc}
\tablewidth{0pt}
\hline
\hline
Parameter & Range & Step Size \\
\hline
T$_{\rm eff}$ [K] & 300$-$950 & 50 \\
log(g) [cgs] & 3.0$-$5.5 & 0.5\\
$[$M/H$]$ & -1$-$1& 0.5\\
C/O & 0.1$-$0.7 & 0.2\\
 & 0.7$-$0.9 & 0.05\\
log(K$_{zz}$) & 2$-$8 & 2\\
\hline
\end{tabular}
\end{table}

\subsubsection{Chemical Trends} \label{sec:chemtrends}

	One of the defining features of a classic retrieval is in its ability to directly constrain atmospheric {\it molecular} abundances from the spectra, free from the a priori assumptions commonly made in self-consistent models. Molecular abundance trends with other properties provide insight into the chemical and physical processes operating in the atmospheres. The retrieved molecular abundances for the ensemble of HST WFC3 late-T and early-Y dwarf are given in Table \ref{tab:abundance}. Here, we focus on how these abundances vary with effective temperature as this is predicted to be the dominant abundance controlling factor through equilibrium chemistry \citep{1999ApJ...512..843B, 2002Icar..155..393L,2007ApJS..168..140S}.
	
	Figure \ref{fig:bigabund} summarizes these trends (red, yellow points) in comparison to predictions from a self-consistent grid model (black curves) and to those derived for the warmer T-dwarfs from Part II (blue points). Our chosen grid model was introduced and validated in \citet{2018AJ....156..133P} and \citet{2018A&A...618A..63B}. We produce a grid of models given T$_{\rm eff}$, log(g), metallicity, and assume radiative-convective thermochemical equilibrium. The molecular abundance curves here are column weighted mixing ratio over the photosphere. 
	
	We find that the H$_2$O and CH$_4$ abundances show a systematic offset between the late-T and early-Y sample. For context we've also plotted column-integrated abundance trends from our grid model, showing that the variation we see between the two samples is largely reproducible by variations in both the C/O ratio and in the metallicity of the system. This falls in line with predictions from equilibrium chemistry where H$_2$O and CH$_4$ remain relatively constant for a given set of elemental abundance assumptions, and that neither molecule should be sensitive to other chemical processes such as vertical disequilibrium mixing \citep{1999ApJ...512..843B,2007ApJS..168..140S}.

	Part II found no systematic trend in the ammonia abundance with effective temperature, despite thermochemical equilibrium predicting a $\sim$0.5 dex increase over the 800-600K temperature range for a given metallicity.  Ammonia is well known to be influenced by vertical mixing at these cool temperatures. Vertical mixing is expected to quench the ammonia abundance to one order of magnitude {\it lower} than the equilibrium abundance over the photospheric layers \citep{2006ApJ...647..552S, 2015ARA&A..53..279M}. 
	
	However, we note that with the addition of our sample, we see a slight trend of increasing ammonia that is largely consistent with thermochemical equilibrium assumptions at a range of metallicities and gravities. Note that we have not included the yellow points in this analysis given the complications with these objects, highlighted in Section \ref{sec:gravity}. Though it is possible that the ammonia in the atmospheres of these objects is being affected by disequilibrium mixing at varying strengths, the ability to test such ideas quickly becomes limited by both the sparse number of retrieved NH$_3$ abundances, and the precision of our retrieval constraints.
    
    \begin{figure}
    \centering
	\includegraphics[width=\linewidth]{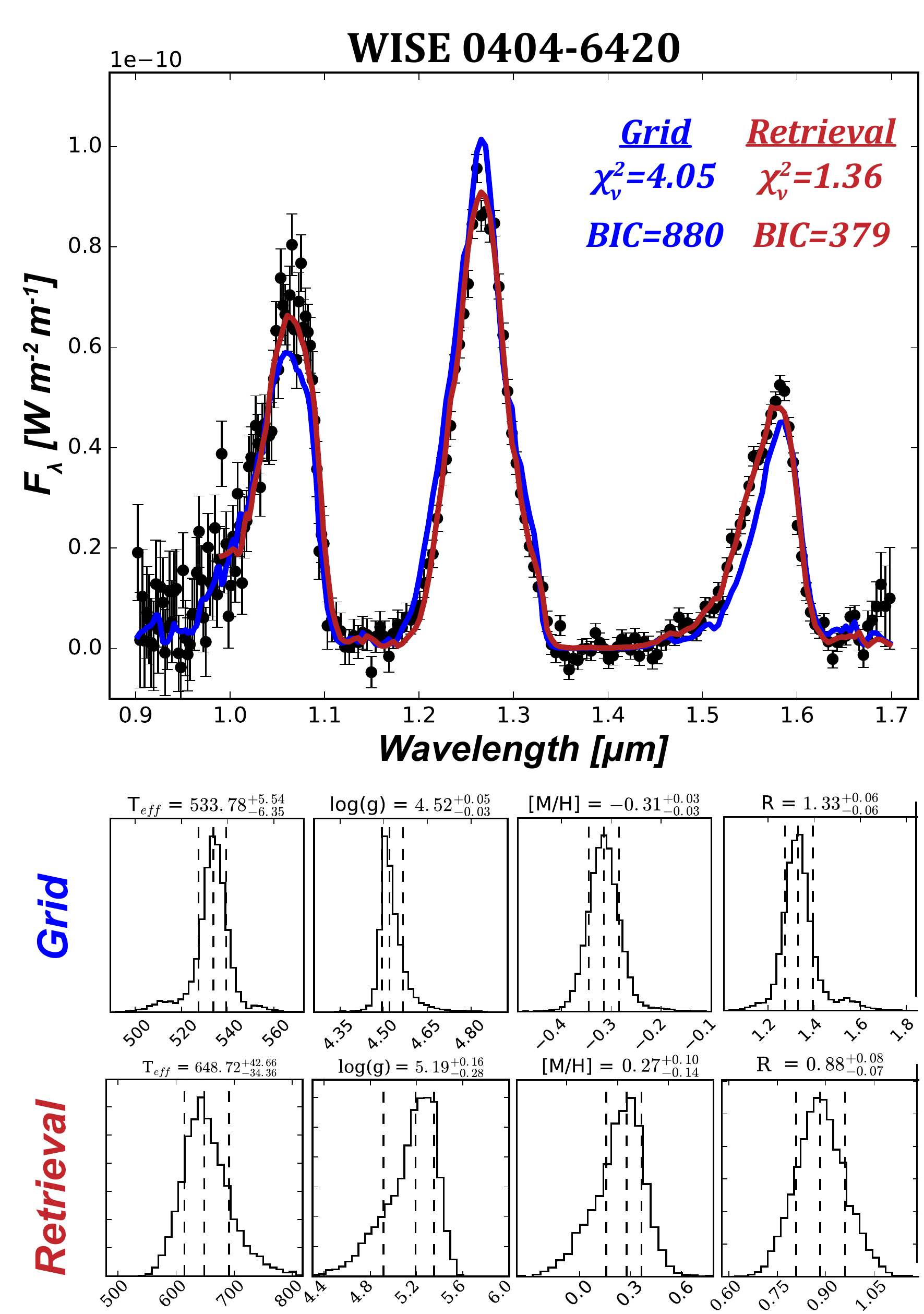}
	\caption{(Top) Best-fit grid-model (blue) and retrieval (red) results for W0404. With only 4 free parameters (T$_{\rm eff}$, log(g), [M/H] and $R$) the grid model struggles to well-fit the entire YJK-band spectra with systematic offsets in each band. Comparing the Bayesian Information Criterion (BIC) between both models suggests the retrieval method is highly preferred. (Bottom) Marginalized posteriors for the relevant free parameters in each model. The poor fit of the grid model often disagrees with the retrieval model and obtains nonphysical constraints.}
    \label{fig:gridfit}
    \end{figure}
    
    A more striking compositional trend, extending far beyond the results in Part II, is that of the alkali metals with temperature. The retrieved Y-dwarf alkali abundances fall off substantially with temperature relative to the warmer T-dwarfs. In all but two cases (W0325,W0335), we only obtain upper limits on the alkali abundances due to the lack of detectability. These results are consistent with predictions from equilibrium rainout chemistry (blue, solid) and strongly disfavor pure equilibrium (blue-dashed, from \citet{2001RvMP...73..719B}). Pure equilibrium permits the existence of aluminum and silicates in the middle atmosphere which achieve equilibrium with the Na and K to form sanidine (KAlSi$_3$O$_8$) and albite (NaAlSi$_3$O$_8$) \citep{2001RvMP...73..719B}, resulting in a rapid depletion of gaseous Na and K at $\sim$1300K. In contrast, rainout rapidly removes aluminum/silicates leaving behind the gas phase alkalies until $\sim$700K where they begin to condense into KCl and Na$_2$S \citep[e.g.][]{2001RvMP...73..719B}. These results are the first to show that a number of indirect lines of evidence for rainout from both pre-computed grid models \citep[e.g][]{2002ApJ...568..335M,2012ApJ...756..172M,2014ApJ...787...78M}, and observations of reddening Y-J colors \citep[e.g.][]{2012ApJ...758...57L,2015ApJ...804...92S} are directly owed to the depletion of Na and K.
    
    We obtain two bound constraints for W0325 and W0335, and only lower limits for cooler targets as the alkalies deplete below retrievable abundances. We note that the one anomalous lower limit at roughly 650K is W1639 whose temperature structure strongly deviates from the typical radiative-convective equilibrium. As a result it is not surprising to find the upper limit for the abundance of this target is systematically shifted from the remainder of our curve. Additionally the results for our three other objects with questionable abundance constraints (W2212, W1405, W1541) still show good agreement with the solar metallicity trend, though this may be a result of only obtaining upper limits for these targets.
    
    Improved SNR and spectral resolution with JWST, particularly at the blue end of the Y band, and near roughly 1.2$\mu m$ where the resonance features for Na and K peak, should allow us to probe cooler objects with far more depleted alkali abundances or uniquely constrain both Na and K independent of each other. In addition, improved NH$_3$ constraints on a larger number of Y-dwarfs may also allow us to directly confirm the presence of vertical disequilibrium mixing in the future. 


\subsection{Grid Model Fitting} \label{sec:grid}

While the retrieval-based approach is useful in its ability to place as little a priori information as possible into the atmospheric model, it is still useful to compare such results against a grid-based model. Grid models incorporate more assumptions and are presumably more self-consistent in that they often treat the atmosphere under radiative-convective-thermochemical equilibrium whereas our retrieval method makes no such assumptions. This is useful in the investigation of both missing model physics within the established grid models, and any possible nonphysical results from the retrieval method as we have seen with our evolutionary parameters.

We use a newly developed grid of self-consistent, cloud-free atmospheric models (Self-consistent CHIMERA, ScCHIMERA) \citep{2018AJ....156..133P,2018A&A...618A..63B}, which utilizes the same underlying radiative transfer and opacity sources as the retrieval forward model. Briefly, the self-consistent model solves for layer mid-point fluxes using the \citet[]{1989JGR....9416287T} two stream source function approach. The model is iterated to radiative equilibrium using the Newton-Raphson method until there is zero net flux divergence throughout the column. Convection is implemented through a mixing length flux \citep[e.g.][]{2015ARA&A..53..279M}. Line-by-line cross-sections are converted to R=100 correlated-K coefficients between 0.3 and 100$\mu$m (using 20 Gauss quadrature points per wavenumber-bin) utilizing the ``resort-rebin" \citep{2017A&A...598A..97A} optical depth approach to speed up efficiency but to maintain accurate flux computations. The converged models are ``post-processed" to an R=1000 (again with correlated-K). These moderate-resolution spectra are then convolved and binned to the data wavelength grid when undergoing fitting. The grid is generated as a function of T$_{\rm eff}$, log(g), [M/H], the C/O ratio, and the vertical eddy diffusion $K_{zz}$ (through the \citet[]{2014ApJ...797...41Z} quench-time scale framework). The grid model parameter ranges and step sizes are given in Table \ref{tab:gridmod}. Using \verb|emcee| and an interpolating function (a variant of Python's \verb|griddata| routine) we fit each object with this 5-dimensional grid, but have also experimented with different subsets of parameters (e.g., fitting for only log(g) and T$_{\rm eff}$ while fixing composition to solar).

Figure \ref{fig:gridfit} shows an example comparison (for W0404) between the grid model solutions and the retrieval solutions. In this specific instance, T$_{\rm eff}$, log(g), [M/H] and the radius-to-distance scaling are the free parameters of the grid with no quenching. From a visual standpoint, there are noticeable differences between the grid model fit and the retrieval fit. The best fitting grid model under fits the Y-band peak and overestimates the J-band peak by  $\sim$10-20\%, as well as the entire blue edge of the H-band. This issue of either overestimating the J-band, underestimating the Y and H bands, or both, is consistent across all of our objects. This result is not unique to our grid model, as previous work using other cloud-free grid models have had similar issues \citep{2015ApJ...804...92S,2017ApJ...842..118L}.

\begin{figure*}
    \centering
	\includegraphics[width=\linewidth]{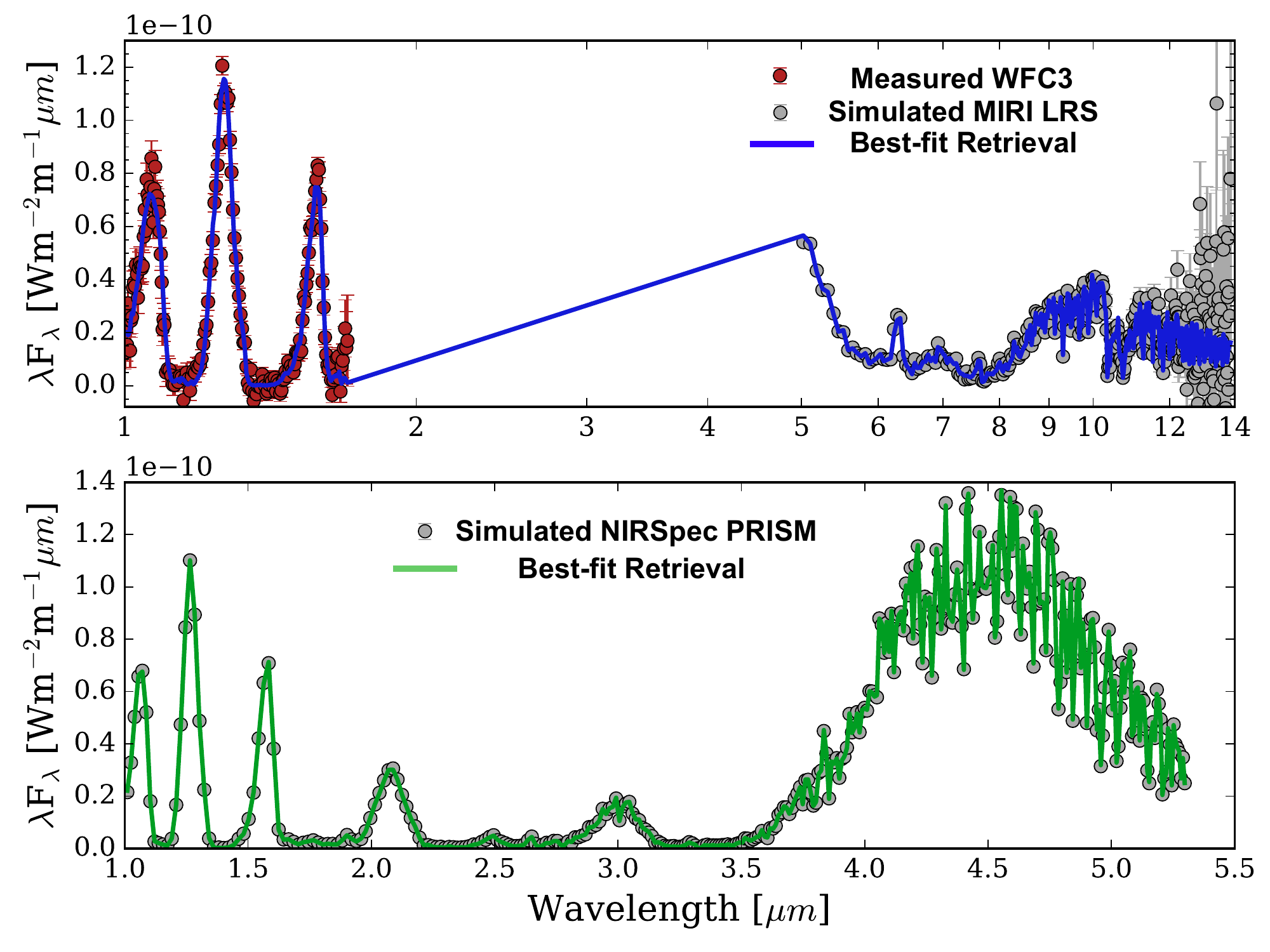}
    \caption{\textit{Top:} Best fit spectrum (blue) to a combined WFC3 observation (1-1.7$\mu m$, red) and JWST MIRI LRS simulation (5-14$\mu m$, grey). \textit{Bottom:} best-fit spectrum (green) to a simulated JWST NIRSpec PRISM spectrum (grey). NIRSpec provides vastly improved SNR (200 vs. 10) for a much shorter exposure time (15mins vs 1hr) when compared to MIRI LRS.}
    \label{fig:jwstspec}
    \end{figure*}

The grid model best fit produces a $\chi^2/N=4.05$ compared to the retrievals $\chi^2/N=1.36$.  We utilize the Bayesian information criterion (BIC) to determine the balance between improved fit and increased parameters and whether the retrieval parameters are indeed justified. The retrieval forward model includes 31 free parameters and 175 data points (we stop at 1$\mu$m due to constraints on the molecular cross-sections) giving a $BIC=379$. The self-consistent grid fit has only 4 free parameters (in this example) and 212 data points (the grid model goes down to 0.9 $\mu$m) resulting in a $BIC=880$. The $\Delta BIC=501$ overwhelmingly favors the retrieval fit according the Jeffery's Scale \citep{10.2307/2291091}. Regardless of the number of free parameters we include in our grid model (including the vertical mixing and carbon-to-oxygen ratio dimensions), we often find similar misfits.
    
\begin{figure*}
	\centering
	\includegraphics[width=\textwidth]{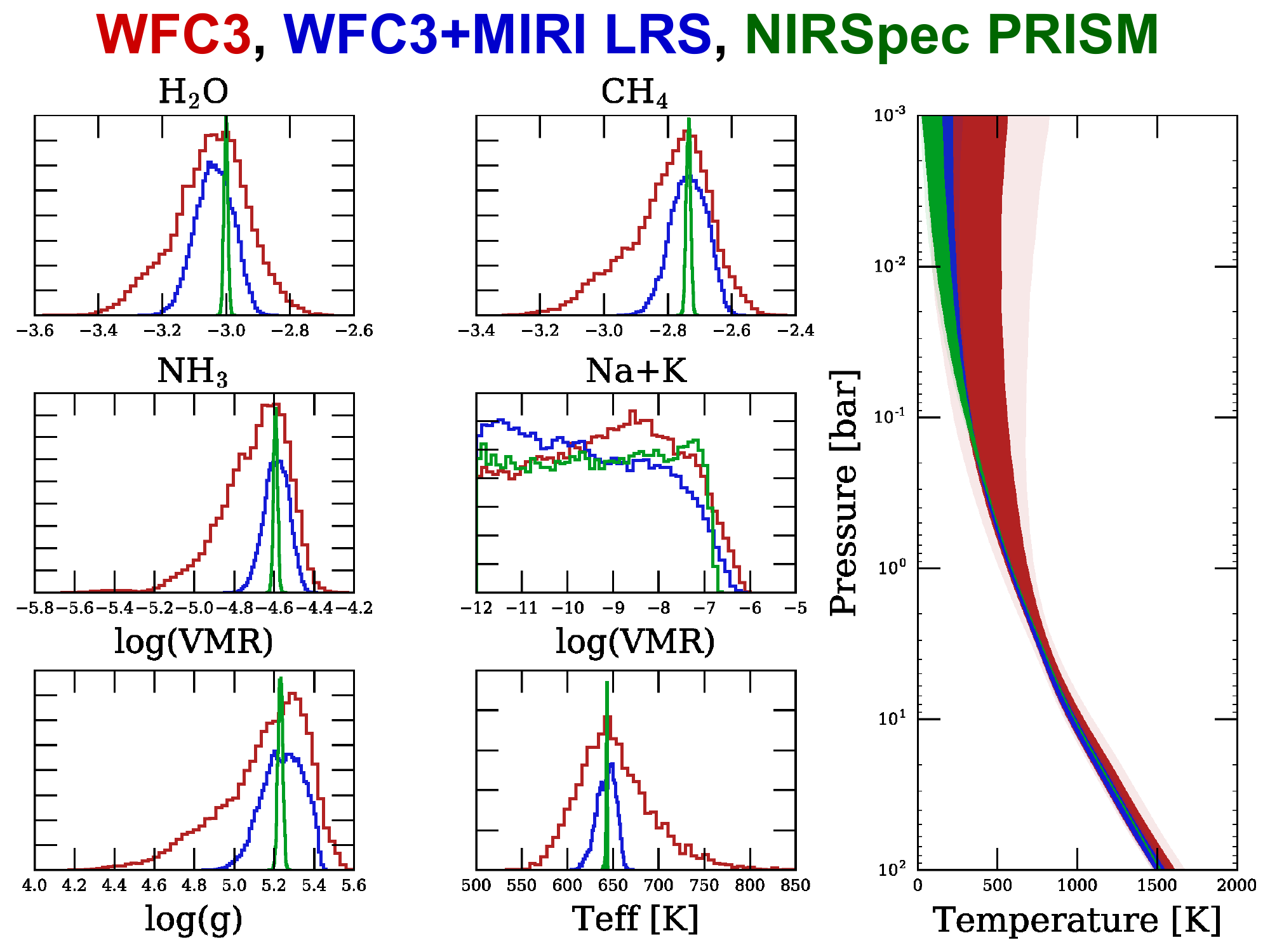}
    \caption{Best-fitting model parameters from our analysis of WFC3 spectra (red). Overlaid are the resulting JWST NIRCam PRISM datapoints and error estimates from the ETC (green). An additional retrieval using a combined WFC3 and MIRI LRS spectrum is also shown for comparison (blue). A NIRCam PRISM observation provides substantially higher precision on molecular abundances and atmospheric structure than a combined WFC3 and MIRI LSR spectrum for about a quarter of JWST exposure time.}
    \label{fig:jwstparams}
	\end{figure*}

Figure \ref{fig:gridfit} also compares the retrieval and grid-model constraints on effective temperature, gravity, metallicity, and radius. We find (consistent amongst our other objects) that the retrieval and grid models often disagree by at least several sigma in almost all model parameters. In the specific example of W0404, the grid model derived effective temperature disagrees with our retrieval result by over 100K, the gravity estimate is inconsistent to almost a full order of magnitude, the metallicity is sub-solar for the grid model yet super-solar for our retrieval, and the radius is inflated in the grid model fit.

For our other targets, the grid model often requires either unphysically high or low: radii, masses, and gravities for typical field brown dwarfs, as well as effective temperatures inconsistent with previously measured spectral types. A full database of all fits, and resulting model parameters, is available at our previously linked Zenodo site. This highlights the need for a retrieval methodology to fully utilize the information content contained in substellar atmospheric spectra in order to accurately characterize both current and future datasets.


\section{JWST Simulation Constraints} \label{sec:jwst}
    
JWST promises to revolutionize our knowledge of brown dwarf atmospheres due to: a vastly improved wavelength coverage across the near and mid infrared, combined with improved SNR and spectral resolution \citep{2009ASSP...10..101M}. Here we take a preliminary look at the potential improvement in our retrieval parameters with JWST for a representative T9 object (W0404).
    
We take the best fitting model to the HST data for W0404 (that is our best-fit model with the parameters specified in Tables 3, 4, and associated figures) and simulate both Near InfraRed Spectrometer (NIRSpec) PRISM and Mid-InfraRed Instrument, Low-Resolution Spectroscopy (MIRI,LRS) observations using the JWST Exposure Time Calculator (ETC) v1.3. The largest 1.6'' slit was chosen for the PRISM/CLEAR configuration and a slitless spectroscopy mode for MIRI LRS were chosen to minimize potential systematic slit-losses from the instrument. We set the integration time to obtain, somewhat arbitrarily, SNR$\approx$200 at the J-band peak within the PRISM mode and SNR's$\approx$10 over MIRI LRS. We found this was achievable with 15 minutes and 1 hour of exposure time on NIRSpec and MIRI respectively.
    
We then applied the same retrieval tools to this simulated data set, under the same exact model assumptions, comparing three cases: WFC3 only (this work), WFC3+MIRI LRS, and NIRSpec PRISM only (Figure \ref{fig:jwstspec}). Figure \ref{fig:jwstparams} summarizes the constraints (red=WFC3 only, blue=WFC3+MIRI, green=NIRSpec only). It is clear that JWST will provide astounding improvements on the molecular abundances, gravity, and temperature profile. For example, we find that the H$_2$O abundance constraint improves from $\pm$0.1dex with WFC3 to roughly $\pm$0.06dex with WFC3 combined with an hour of MIRI LRS integration time, and better than $\pm$10\% for only 15 minutes of NIRSpec integration time. These extremely tight constraints approach the precision of remote solar-system quality science on brown dwarfs, and speak to the utility of JWST to well-characterize nearby substellar atmospheres in the future.

One caveat here is that this analysis makes the assumption that our model that best-fits the YJH bands of WFC3 is an accurate representation of the object's spectra at both longer wavelengths and higher spectral resolutions. Additionally, such an analysis does not account for any potential systematics, currently known or unknown, that will impact the future performance of JWST that are not properly accounted for in the JWST ETC. These systematic biases between instruments, or within JWST itself, will lower the precision of constraints shown here. Despite these limitations, such an analysis provides an initial first step in understanding how well JWST will be able to constrain atmospheric properties on cool brown dwarfs.

\section{Conclusions} \label{sec:conclusions}

    We have extended the work of previous investigations using a well-vetted atmospheric retrieval approach into the cooler Y dwarf spectral class. This is done by comparing our model to a set of uniformly reduced, low-resolution WFC3 measurements for an ensemble of late-T and early-Y dwarfs. Such a methodology has provided the first direct constraints on the chemical composition of cool Y dwarfs and provides a foundational dataset that can be compared to future low-mass characterization work. Our main scientific results are as follows:

	\begin{enumerate}
	\item We are able to well-fit our ensemble of late-T and early-Y dwarfs with our retrieval model across the YJH bands as shown in Figure \ref{fig:bigspec}, Section \ref{sec:results}. We find no systematic deviations from the data in our residuals. This is in contrast to typical grid modeling efforts which often miss key spectral features of these cooler objects.
    \item Overall the retrieved temperature structures are consistent with radiative-convective equilibrium except in the marked case of W1639 whose peculiar Y band structure may be indicative of a non-radiative-convective equilibrium structure in Figure \ref{fig:bigspec}, Section \ref{sec:temp}. However, inconsistencies in derived evolutionary parameters may also indicate our model is not well adapted to explain the odd Y band structure.
    \item For most of our objects, we obtain mass estimates that are consistent with field-age brown dwarfs but systematically smaller radii, and higher gravities than allowable with evolutionary models (see Figure \ref{fig:logg_teff}, Section \ref{sec:gravity}). We attempted a myriad of tests on both the observational data and our retrieval model to discover the cause of this deviation. Using a distance estimate from another parallax program, we found that W2056's anomalously high gravity could be explained by a systematic bias in the distance estimate. If the distances are all systematically underestimated, this would explain both our high gravities and lower radii for the majority of our objects. More importantly this indicates how sensitive our retrieved results can be to small changes in distance estimates. The coldest Y dwarfs, W1405 and W1541's results are speculative at best given the retrieved masses and radii are inconsistent with known limits for field-age brown dwarfs.
    \item We obtain for the first time, direct, bound constraints or upper limits on H$_2$O, CH$_4$, CO, CO$_2$, H$_2$S, NH$_3$, and Na+K for an ensemble of cool Y dwarfs (see Section \ref{sec:comp}). From these measurements we drive preliminary C/O and metallicity estimates that, when oxygen sequestering via chemical rainout of silicates is taken into account, are broadly consistent with the local FGK stellar population, albeit at slightly enhanced metallicity.
    \item From these measurements we investigate chemical trends with T$_{\rm eff}$ which are diagnostic of the chemical mechanisms at work in the atmospheres of brown dwarfs. We find that H$_2$O and CH$_4$ are consistent with expected chemical equilibrium predictions and are not subject to either chemical rainout or vertical disequilibrium mixing. NH$_3$ may show a tentative trend with either pure chemical equilibrium or disequilibrium vertical mixing. Improved constraints from JWST would be more diagnostic of this trend and may be able to constrain the strength of mixing. Finally, Na+K shows a trend consistent with both chemical rainout and the results in Part II, as opposed to pure chemical equilibrium. This result confirms that the blue shift in the Y-J color photometry across the T/Y boundary is owed to the depletion of alkali metals.
    \item We make predictions for future JWST observations for cool late-T and early-Y dwarf targets. We find that NIRSpec offers the best observing mode in order to do high-precision abundance measurements on near-by brown dwarfs, approaching that of current bulk solar-system quality measurements. Such high precision abundance measurements provide useful diagnostic for future modeling efforts to understand cool, substellar atmospheres.
	\end{enumerate}

\acknowledgments

We thank Roxana Lupu and Richard Freedman for continued development and support of their extensive opacity database which makes much of this work possible. We also thank Dan Foreman-Mackey for his publicly available EMCEE code and the useful plotting routine \verb|corner.py|. We thank Adam Schenider, Mark Marley, Jennifer Patience, Laura Kreidberg, Michael Cushing, Jackey Faherty, Trent Dupuy, and Wanda Feng for the many comments, discussions, and tools which have improved this work. This research has benefited from the Y Dwarf Compendium maintained by Michael Cushing at \url{https://sites.google.com/view/ydwarfcompendium/}. This material is based upon work supported by the National Science Foundation under Grant No. AST-1615220.

\software{python2.7, matplotlib, CHIMERA \citep{2013ApJ...775..137L}, ScCHIMERA \citep{2018AJ....156..133P,2018A&A...618A..63B}, emcee \citep{2013PASP..125..306F}, corner.py}

\bibliography{bibliography}

\end{document}